\documentclass[twocolumn,doublespacing,noshowpacs,showkeys,preprintnumbers,nofootinbib,amsmath,amssymb]{revtex4}
\usepackage[english]{babel}
\usepackage{epsfig,amsfonts}
\usepackage{bm}

\newcommand\De{\Delta}

\newcommand\Lam{\Lambda}

\newcommand\half{{\frac{1}{2}}}
\newcommand\tralf{{\frac{3}{2}}}

\newcommand\lam{\lambda}

\begin{document}

\title{Dibaryon resonances and short-range $\bm{NN}$ interaction}
\author{V.I. Kukulin}
\email{Deceased}
\author{V.N. Pomerantsev}
\email{pomeran@nucl-th.sinp.msu.ru}
\author{O.A. Rubtsova }%
\email{rubtsova-olga@yandex.ru}
\author{M.N. Platonova }%
\email{platonova@nucl-th.sinp.msu.ru}
\author{I.T. Obukhovsky }%
\email{obukh@nucl-th.sinp.msu.ru}

\affiliation{%
Skobeltsyn Institute of Nuclear Physics, Lomonosov Moscow State
University, Leninskie Gory 1/2, 119991 Moscow, Russia}

\date{\today}

\begin{abstract}
The dibaryon concept for the nuclear force is presented, assuming that the main
attraction between the nucleons at medium distances is determined by the $s$-channel
exchange of an intermediate six-quark (dibaryon) state. To construct the respective $NN$ interaction
model, a microscopic six-quark description
of the $NN$ system is used, in which symmetry aspects play a special role. It is shown that the $NN$ interaction in all important partial waves can be described properly by a superposition of the long-range $t$-channel one-pion exchange and the $s$-channel exchange by an intermediate dibaryon. The developed model gives a good description of both elastic phase shifts and inelasticities of $NN$ scattering in all $S$, $P$, $D$ and $F$ partial waves at energies from zero to 600–800 MeV and even higher.
The parameters of the intermediate six-quark states corresponding to the best fit of $NN$ scattering data are found to be consistent with the parameters of the known dibaryon resonances in those $NN$ partial configurations where their existence has been experimentally confirmed. Predictions for new dibaryon states are given as well.
\end{abstract}


\thanks{The work has been partially supported by
RFBR, grants Nos. 19-02-00011 and 19-02-00014. M.N.P. also appreciates support from the
Foundation for the Advancement of Theoretical Physics and
Mathematics ``BASIS''.}

\maketitle
\section{Introduction}
\label{sec_intro}

It is now generally accepted that fundamental quantum chromodynamics (QCD) is the basis of nuclear physics and determines its basic phenomena and laws, such as the nature of nuclear forces, the specific structure of nuclei, the observed features of nuclear reactions, etc. However, the direct connection between QCD and nuclear physics is hidden behind the complicated dynamics of quarks and gluons in the nonperturbative region, where a highly nontrivial relation between the completely different degrees of freedom of QCD and nuclear physics takes place. While QCD operates with quarks and gluons, in traditional nuclear physics we deal with nucleons, mesons and nucleon isobars.

Despite the fact that there are a lot of QCD-motivated models, which can
reproduce the basic properties of baryons and mesons, still one has no
quantitative approaches to connect
the observable properties of $NN$ and $3N$ interactions, like $NN$
scattering phase shifts and inelasticities, the binding energies of two-
and few-nucleon systems, etc., with the underlying properties of QCD. There
are a number of quark models which describe $NN$
elastic scattering (see, e.g., the reviews~\cite{Q1,Q2}), however, there are only a few (if any) quark-model
approaches to the description of both elastic and inelastic $NN$ collisions
above the pion-production threshold. Moreover, the quark models developed to
date do not provide a quantitative description of nuclear phenomena.

Thus, the constituent quark model (CQM), which uses the QCD-motivated quark-quark interaction, quite successfully describes the phase shifts of $NN$ elastic scattering below the pion production threshold (and also the deuteron properties) based on calculations made within the resonating group method (RGM). In this case, the mixing of quark configurations and the connection of the $NN$ channel with other cluster channels ($N\Delta$, $\Delta\Delta$, $NN^*$) play a crucial role~\cite{Q2}. Without taking these effects into account, the description of $S$ and $D$ waves would not be so satisfactory. In addition, an adequate description of the triplet $P$ and $F$ waves has not yet been obtained. No RGM calculations were performed above the pion production threshold, and it is clear that with increasing energy the role of virtual clusters will increase as well and they will make a nontrivial contribution to the production of real baryons. It is unlikely that calculations of such inelastic processes could be realized starting from the pair $qq$ interactions. On the contrary, the model proposed below is effective in calculating both elastic and inelastic phase shifts in an energy range from zero to about 1 GeV, and instead of a set of $3q-3q$ virtual cluster channels, it uses only one additional channel with a $6q$ dibaryon, which has quantum numbers, mass and width of the actually observed (or predicted) resonances.



In fact, the six-quark (or dibaryon) resonances can serve as an effective tool to relate
the world of QCD with that of two- and few-nucleon dynamics. The dibaryon resonances,
being essentially multi-quark objects, reflect in their structure the very
complicated dynamics of quarks and gluons, on the one hand, and,
considered as dinucleon systems, reflect the basic properties of the short-range
$NN$ interaction, on the other hand. Hence, the dibaryon (and, more generally, multi-baryon) resonances
can be considered as appropriate
effective degrees of freedom to describe the nuclear-physics phenomena which
involve the short-range $NN$ interaction~\cite{Kuk-YAF11}.
It should be emphasized here that dibaryon resonances can give a lot for understanding
the short-range $NN$, $N\De$, $\De\De$, etc., forces not only due to their structure
but, first of all, because they are specific relatively long-lived states in which
six-quark dynamics should be manifested. It should be
noted also that in the majority of the previous experimental and theoretical
studies, the non-trivial hexaquark states have been treated as some exotics, like penta- or
tetraquarks, which are not related directly to the basic mechanism of the $NN$ interaction.

The existence and main properties of some dibaryon resonances have now been
reliably confirmed in the numerous experimental and theoretical studies (see the
recent reviews~\cite{Clem17,Clem21,Gal16}). It should be noted that the history of
dibaryon resonances was very dramatic, from the initial enthusiasm through
years of skepticism or even complete rejection until the final discovery in a
series of precise high-statistics experiments made by several international
collaborations (CELSIUS/WASA, WASA-at-COSY, ANKE-COSY, and others).

The dibaryon concept for the nuclear force based on an idea of the $s$-channel
exchange dominance, originally proposed in Ref.~\cite{PIYAF}, occurred to be very
fruitful and resulted in the construction of the dibaryon model (initially called the ``dressed-bag model'') for the $NN$ interaction~\cite{JPhys2001,IntJModPhys2002}. To keep the connection with the conventional meson-exchange ideas, the long-range part of the interaction was treated via the one-pion-exchange potential (OPEP). At the same time, the traditional $t$-channel multi-meson exchanges at short $NN$ distances were replaced by the $s$-channel mechanism corresponding to the exchange of the dibaryon resonance (the $6q$ bag dressed with meson fields) between the interacting nucleons. Such a replacement looks quite natural in the two-nucleon overlap region and implements the duality principle for $NN$ scattering (see, e.g., Ref.~\cite{dual}). Moreover, this allows one to overcome a number of difficulties and inconsistencies in the coupling constants, cut-off parameters, etc., which persist in the meson-exchange approaches (for a detailed discussion on this issue, see, e.g., Ref.~\cite{PAN13}). The initial version of the dibaryon model provided a very
good description for $NN$ elastic phase shifts in the lowest partial waves at
energies up to 600 MeV (lab.) as well as the deuteron properties~\cite{JPhys2001,IntJModPhys2002}. A review of successes and consequences of the dressed-bag model can be found in Ref.~\cite{AnnPhys2010}.
However, the parameters of the resonance poles obtained in this initial model have never been compared with the parameters of dibaryon resonances deduced from experimental data by the partial-wave analysis (PWA) or phenomenological models.

Recently, we extended the dibaryon model~\cite{JPhys2001,IntJModPhys2002} to higher partial waves and took into account inelastic processes (see Refs.~\cite{YAF19,PLB20,EPJA20,PRD20}). This new version of the model makes it possible to describe both elastic and inelastic $NN$ scattering in a wide energy range far above the pion production threshold as well as to reproduce (or predict) the empirical parameters of dibaryon resonances.

The main goal of the present paper is to demonstrate that the $NN$ interaction in all important partial waves can be described properly by a superposition of the long-range $t$-channel one-pion exchange and the $s$-channel exchange by an intermediate dibaryon state.
Thus, we argue that the $s$-channel mechanism proposed not only looks more natural but also can effectively replace the conventional $t$-channel multi-meson exchange at short distances (in the two-nucleon overlap region).
We present here the modified version of the dibaryon model and make a comprehensive analysis of $NN$ scattering in the framework of this model including a number of partial-wave configurations not considered in the previous works. We should emphasize that our purpose is not to prove the existence of dibaryon resonances, but to use them for construction of some alternative picture of the $NN$ interaction and then compare their parameters with those deduced from experimental data. The success of this picture does not mean that one should reject the huge progress achieved within the conventional approaches. The short-range $NN$ interaction can be actually described in different ways, but the QCD-motivated approach presented here has a wider range of applicability since it is free of some limitations inherent to the existing quark-model or meson-exchange treatments.

The structure of the paper is as follows.
In Sec.~\ref{sec_status} we outline the modern experimental status of dibaryon
resonances and their possible theoretical interpretation. In
Sec.~\ref{sec_quark}, we describe the basic assumptions of the dibaryon concept
and the tools needed for constructing the model for the $NN$ interaction. Here, we introduce the
two-channel formalism for the $NN$ system with an additional internal channel
corresponding to the quark degrees of freedom and present the basic results of
the $6q$ microscopic consideration for the $NN$ system. In particular, we
demonstrate here that all possible $6q$ states can be divided into states with a
$6q$ bag structure with no leading hadronic configuration and two-cluster states
with the dominating $NN$, $N\Delta$, $\Delta\Delta$, etc., configurations. In
Sec.~\ref{sec_effham} we derive the effective Hamiltonian of the dibaryon model
using a simple one-pole approximation for the internal channel resolvent. In
Sec.~\ref{sec_recent}, the latest version of the model with account for inelastic processes is described and the
results of calculations of phase shifts, inelasticities, and resonance
parameters for the particular $NN$ partial waves are presented. In
Sec.~\ref{sec_concl}, we summarize our results and conclude. In the Appendices
\ref{formfactor} and \ref{dressing}, for the readers' convenience, we
briefly reproduce the quark-model calculations of the transition form factors
and vertex functions for the effective Hamiltonian, following Ref.~\cite{IntJModPhys2002}.

\section{Modern status and structure of dibaryons}
\label{sec_status}

In this section, we give a brief review of the modern status of dibaryon resonances, from the experimental and theoretical point of view.
This is needed to substantiate the dibaryon concept for the $NN$ interaction described in the next sections.
For a more comprehensive review, see Refs.~\cite{Clem17,Clem21,Gal16}.
\subsection{Modern status of dibaryon resonances}
\subsubsection{Resonances predicted by Dyson and Xuong}
The first theoretical prediction for the existence of hexaquark (or dibaryon) resonances was done by Dyson and Xuong \cite{Dyson64} as early as in 1964, i.e., very soon after a pioneer work of Gell-Mann about quarks.
By using the SU(6) symmetry, the authors~\cite{Dyson64} predicted three pairs of low-lying non-strange dibaryon
states near $NN$, $N\Delta$ and $\Delta\Delta$ thresholds. Denoted by $\mathcal{D}_{TJ}$, where $T$ and $J$ mean the isospin and total angular momentum, there were $\mathcal{D}_{01}$ (the deuteron) and $\mathcal{D}_{10}$ (the singlet deuteron), $\mathcal{D}_{12}$ and $\mathcal{D}_{21}$, and $\mathcal{D}_{03}$ and $\mathcal{D}_{30}$ states. Based on the simple SU(6) mass formula and using the known masses of the first two trivial dibaryons, Dyson and Xuong predicted the masses of other four states. Five out of the above six dibaryons have now been
confirmed by experiments, which revealed surprisingly good agreement of the observed dibaryon masses with the above predictions.

In fact, the resonance peak located slightly below the $N\Delta$ threshold and having the width of about the $\Delta$ width was observed well before the prediction of Dyson and Xuong in the experiments made by the group of Meshcheryakov et al. at the Dubna Synchrocyclotron~\cite{Meshcher} on the reaction $\pi^+ d \to pp$. The follow-up PWA~\cite{Hoshiz,Arndt,Strak} confirmed the $\mathcal{D}_{12}$ dibaryon resonance in the reaction $\pi^+ d \to pp$ and revealed it also in $pp$ and $\pi d$ elastic scattering. The resonance pole corresponding to the average mass of about 2160 MeV and width of about 120 MeV was also found in the recent Faddeev calculations of the $\pi NN$ system~\cite{Gal14}. Theoretically, it can be interpreted as an $N\Delta$ molecular-like state, though some admixture of hexaquark components is not excluded as well.

Some indication of the resonance $\mathcal{D}_{03}$ (denoted also by $d^*(2380)$) was found already in the old experiments~\cite{Kamae77} on the reaction $np \to d\gamma$. However, just the recent series of high-statistics measurements made first by the CELSIUS-WASA and then by the WASA-at-COSY Collaborations~\cite{Bash09,Adl1113,AD14} on $pn$-induced double-pion production and $np$ elastic scattering left no doubt in the existence of the $T(J^P)=0(3^+)$ dibaryon state with the mass 2380 MeV (i.e., 80 MeV below the $\Delta\Delta$ threshold) and rather narrow width of about 70 MeV. The $d^*(2380)$ resonance was also confirmed in the deuteron photodisintegration measurements by the A2 Collaboration at MAMI~\cite{Bash-em}. The resonance pole corresponding to the $d^*(2380)$ state was found in the PWA~\cite{AD14} and confirmed by the Faddeev calculations of the $\pi N\Delta$ system~\cite{Gal14}. The subsequent quark-model calculations (e.g.,~\cite{BBC13,Huang}) explained the observed mass and width of the dibaryon as being due to the dominance of the six-quark hidden-color (CC) components in this state. For now, it appears to be the only known dibaryon state having predominantly hexaquark (i.e., not molecular-like) structure~\cite{Clem21}. One should note also a recent attempt to explain the observed properties of interacting $N\Delta$ and $\Delta\Delta$ systems without need for any six-quark states, based on the coupled-channel calculation with account of the Fermi motion in these systems~\cite{Nisk17}. Such a consideration, however, gives too narrow widths of the observed $\mathcal{D}_{12}$ and $\mathcal{D}_{03}$ states and does not agree with some of the experimental mass distributions~\cite{Clem21}.

Two resonances with the mirrored quantum numbers, i.e., $\mathcal{D}_{21}$ and $\mathcal{D}_{30}$, have been actively searched for in the recent WASA-at-COSY experiments on the reactions $pp \to pp \pi^+\pi^-$ and $pp \to pp \pi^+\pi^+\pi^-\pi^-$, respectively. The evidence for the first resonance with the mass and width very close to those of the $\mathcal{D}_{12}$ state has been actually found~\cite{Adl18d21}. This resonance has also been predicted in the Faddeev calculations~\cite{Gal14}. For the $\mathcal{D}_{30}$ state, only upper limits have been found so far~\cite{Adl16d30}, so, this dibaryon state should be investigated further.
\subsubsection{Other isovector resonances}
In the late 1970s, the whole series of isovector dibaryon resonances in the channels with $L=J$ ($^1D_2$, $^3F_3$, $^1G_4$, $^3H_5$, etc.) were discovered in double-polarized $\vec{p}+\vec{p}$ scattering (i.e., the spin-polarized
beam was scattered on the spin-polarized target)~\cite{Auer1,Auer2}. The subsequent analysis~\cite{MacGregor}
showed that these diproton resonances can be arranged to lie on a straight-line
trajectory in the plane $L(L+1) - M_{D}$, where $L$ is the relative orbital angular momentum in the $pp$ system.
This straight-line trajectory was interpreted~\cite{MacGregor} as evidence of the rotational
nature of these diproton resonances very similar to the nuclear rotational bands. Soon
after that, in 1980, the Nijmegen group (Mulders et al.) suggested the $q^4-q^2$ string-like
model~\cite{Nijm} for the six-quark states, which was generalized later by the ITEP group (Kondratyuk et al.)~\cite{ITEP}
with incorporation of the relativistic treatment and spin-orbit splitting in the $6q$
system. In terms of the Nijmegen--ITEP $q^4-q^2$ model, the above series of isovector dibaryons
corresponds to the rotation of the
two-cluster system with the color string connecting the $q^4$ and $q^2$ quark clusters
on its ends.

The direct extrapolation of the trajectory for $L=0$ and $L=1$
gives the $^1S_0$ dibaryon shifted upwards by 145 MeV. This shift might be explained by the intermediate $\sigma$-meson production in the
spin-singlet $^1S_0$ channel, which lowers the mass of the dibaryon (see Sec.~\ref{ITEP}).
The $L=1$ dibaryon should have the quantum numbers $^3P_1$ and the mass $\sim 2060$~MeV. So, this dibaryon could be identified by its mass with the $d'$ resonance predicted in Ref.~\cite{ITEP}. The corresponding resonance peak was observed in the exclusive measurements of the $pp \to pp \pi^+\pi^-$ reaction~\cite{dprime} but has not been confirmed by the succeeding experiments~\cite{dprime1}. In fact, no clear signal of a dibaryon state in the $^3P_1$ channel has been detected to date. However, such a state (with a mass of about 2180 MeV) was found in the PWA~\cite{Strak}. The recent calculations~\cite{PRD20} within the dibaryon model have also shown that the existence of the $^3P_1$ dibaryon resonance with the mass of about 2200 MeV does not contradict the SAID PWA data~\cite{SAID}. However, these data are not sensitive enough to deduce the mass and width of the resonance unambiguously, so, the question about the existence of the $^3P_1$ resonance is still open.

On the other hand, the dibaryon resonances in the $^3P_0$ and $^3P_2$ channels with the mass of about 2200 MeV have been clearly found in the recent experiment of the ANKE-COSY Collaboration~\cite{Komarov16}. The authors~\cite{Komarov16} studied the reaction $pp \to (pp)_S \pi^0$, where $(pp)_S$ is a diproton in the near-threshold $^1S_0$ state. This reaction is complimentary to the reaction $pp \to d \pi^+$, where the isovector dibaryons $^1D_2$, $^3F_3$ and $^3P_2$ play an important role~\cite{NPA2016,PRD2016}. The $^1D_2p$ and $^3F_3d$ transitions which dominate the reaction with the final deuteron are excluded by selection rules in the case of the final diproton. So, the largest contribution here is given by the amplitudes $^3P_0s$ and $^3P_2d$, both of which exhibit the pronounced resonance behavior. While the $^3P_2$ resonance was known previously from the PWA~\cite{Arndt,Strak}, the $^3P_0$ one has been found in Ref.~\cite{Komarov16} for the first time.

\subsubsection{Dibaryon masses and nucleon resonance thresholds}
\label{sec-thr}
Hadron and nuclear physics tell us that bound or resonance states generally appear near the thresholds. It seems true for the known dibaryon resonances as well. As was pointed out in Ref.~\cite{Strak91}, there is a clustering effect for the isovector $^1D_2$, $^3F_3$, $^1G_4$, etc., states, as their masses are close to each other and to the $N\Delta$ threshold. Moreover, these states lie very close to the
thresholds in the respective partial waves, when the orbital angular momentum is taken into account~\cite{Nisk20}.
The $P$-wave states $^3P_0$ and $^3P_2$ found in Ref.~\cite{Komarov16} lie also near the $N\Delta$ threshold.
 At the same time, the isoscalar $d^*(2380)$ state is located rather close to the $\Delta\Delta$ threshold, while the ``trivial'' $S$-wave states $^3S_1$ (deuteron) and $^1S_0$ (singlet deuteron) lie near the $NN$ threshold. Recently, two more dibaryon states near the $NN^*(1440)$ threshold have been found both in the WASA-at-COSY experiments on single- and double-pion production and theoretical calculations of $NN$ elastic scattering in $S$ waves~\cite{EPJA20}\footnote{The recent analysis has shown the dibaryon state found in the isoscalar single-pion production to be related predominantly to the $^1P_1$ rather than $^3S_1$ state~\cite{Clem20}. This situation might be similar to that with several dibaryon states sitting on top of each other near the $N\Delta$ threshold.}.

Besides that, two new isoscalar dibaryons at 2.47 and 2.63 GeV have been found in the recent measurements of double-pion photoproduction on the deuteron at ELPH (Tohoku)~\cite{Ishikawa19}. The positions of these states correspond to the second and third nucleon resonance regions, respectively. The special kinematic constraints of the experiments~\cite{Ishikawa19} made it possible to separate the dibaryon contributions from those of the nucleon resonances. The same experiments have also confirmed the $\mathcal{D}_{03}$ and $\mathcal{D}_{12}$ resonances. The results of Ref.~\cite{Ishikawa19} have been confirmed by an independent experiment at ELSA (Bonn)~\cite{Jude22}, except for the parameters of the $\mathcal{D}_{12}$ resonance which occurred to be rather low.
One more isoscalar resonance has been found by the Tohoku group in the $\gamma d \to d \eta$ reaction~\cite{Ishikawa21}. This resonance with a mass of about 2.43 GeV and a narrow width of only 34 MeV lies near the $NN^*(1535)$ and $d\eta$ thresholds. A similar dibaryon state has been announced also by the ANKE-COSY Collaboration in $pn \to dX$ around the $d\eta$ threshold~\cite{Tsirkov19}. Very recently, an isovector resonance with a mass of about 2.65 GeV and width of 260 MeV has been reported in the analysis of the ANKE-COSY experiment on $pp \to (pp)_S \pi^0$~\cite{Tsirkov22}. Preliminary theoretical calculations suggest this state to have predominantly the $\Delta N^*(1440)$ structure.

Thus, dibaryon resonances have been discovered to date in almost all basic $NN$ partial channels. Besides that, there are indications (or evidences) of some states uncoupled from the $NN$ channel, which can be manifested in the $NN\pi$, $NN\pi\pi$, etc., systems. All known dibaryon states are located near the respective di-hadron thresholds. These findings are very inspiring for searching new near-threshold dibaryons and developing the classification of these states to shed light on their microscopic structure.

\subsection{Two-cluster $q^4-q^2$ model for dibaryon states and $\sigma$-meson emission}
\label{ITEP}

Among many theoretical models for dibaryon states, the closest one to our consideration is the Nijmegen--ITEP $q^4-q^2$ model~\cite{Nijm,ITEP}. We started our six-quark studies from the other edge, i.e., using the quark shell-model representation (see details in Refs.~\cite{JPhys2001,IntJModPhys2002}) to describe the bag-like multi-quark states. Nevertheless, we found that two above completely different pictures, i.e., the two-cluster $4q-2q$ and $6q$ shell-model representations, can be rewritten in a unified form. In fact, the leading shell-model $6q$ configuration $|s^4p^2[42]L=0,2;ST\rangle$ (written in a single-particle representation) can be transformed into the two-cluster $4q-2q$ form using the standard Talmi--Moshinsky transformation for the harmonic oscillator functions (h.o.f.):
\begin{equation}
|s^4p^2[42]LST\rangle \Rightarrow |s^4[4]L_0S_0T_0,s^2[2]lst(2\hbar\omega)L=0,2ST\rangle,
\end{equation}
where the tetraquark $|s^4[4]L_0S_0T_0\rangle$ and diquark $|s^2[2]lst\rangle$ have the $2\hbar\omega$ relative-motion wavefunction and mutual orbital momenta $L=0,2$ allowable for a two-quanta excitation of the h.o.f. The most low-lying six-quark configurations correspond to the tetraquark having $L_0=0$, $S_0T_0=01$ or 10 and the diquark being in a scalar ($lst=000$) or axial ($lst=011$) states~\cite{ITEP}. Two quark clusters are assumed to be connected by a color string with a $2\hbar\omega$ excitation energy. It can be interpreted as a $2\hbar\omega$-vibration or a $D$-wave rotation of the string.
Thus, it turns out that the quark-cluster model suggested by the Nijmegen and ITEP groups and our shell-model picture can be transformed into each other and interpreted in a unified way.

\begin{figure}[h!] \centering\epsfig{file=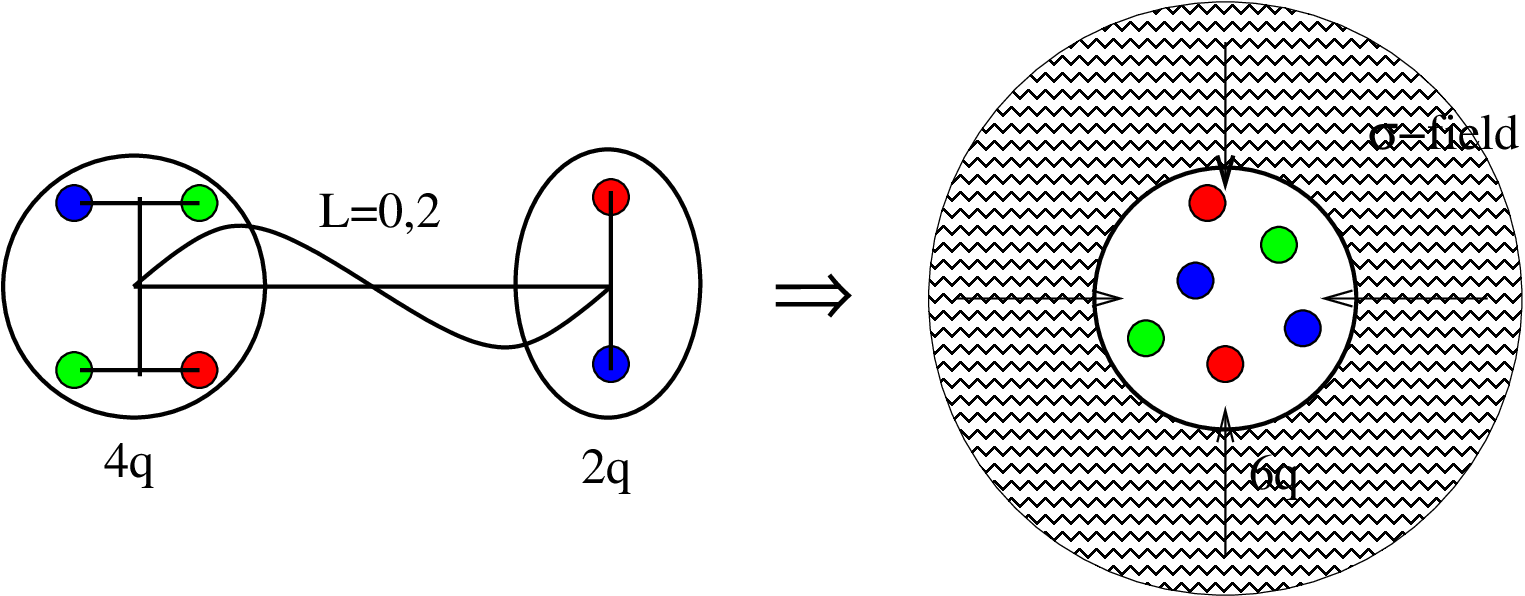,width=\columnwidth}
\caption{\label{4q2q} Illustration for the transition of the $2\hbar\omega$-excited $6q$ state into the unexcited fully symmetric configuration $|s^6[6]+\sigma,L\rangle$ by emission of a scalar $\sigma$-meson from the excited color string.}
\end{figure}

In turn, the $2\hbar\omega$-excited string can emit a scalar $\sigma$-meson and thus, the excited two-cluster $4q-2q$ state can transit into an unexcited bag-like configuration $|s^6[6]+\sigma,L\rangle$ with conservation of the orbital momentum $L$ (see Fig.~\ref{4q2q}). In the quark shell-model representation, it corresponds to the transition of two $p$-shell quarks from the $p$ to $s$ orbit with the simultaneous emission of two tightly correlated $s$-wave pions. For instance, in $S$ waves we have:
\begin{equation}
|s^4p^2[42]_xLST\rangle \to |s^6[6] +
\sigma(l_\sigma=0)\rangle.
\end{equation}

Thus, we identify the specific mechanism of the $\sigma$-meson emission from the excited dibaryons
with the $\sigma$ emission from the excited color string. Such a type of string transition accompanied by the two-pion emission appears to take place in hadronic processes, like the huge $2\pi^0$ production in the scalar mode in
high-energy $pp$ collisions~\cite{Alde97}, the $2\pi$-decay of the Roper resonance, etc.
Besides that, we have shown recently that the emission of the $\sigma$ meson from the intermediate dibaryon state can explain the long-term near-threshold anomaly (the so-called ABC effect) in two-pion production in $pn$, $pd$, etc., collisions at intermediate energies~\cite{PRC2013,PRD2021}.

It is very interesting also to identify this string de-excitation mechanism with the $\sigma$-meson emission
via a monopole transition in the spectra of charmonium and bottomonium:
$$\Psi(2s)\to \Psi(1s)+2\pi^0,\; \Psi(3s)\to \Psi(2s)+2\pi^0, \dots$$
$$\Upsilon(2s)\to \Upsilon(1s)+2\pi^0,\; \Upsilon(3s)\to \Upsilon(2s)+2\pi^0, \dots$$
It is well known that these monopole transitions are associated with
de-excitation of the string connecting $q$ and $\bar q$ in quarkonia~\cite{Eichten}.
Thus, one can suggest, in particular, that two-pion production in high- or intermediate-energy $NN$ collisions and
the monopole transitions in the quarkonia spectra have the unified nature related to
de-excitation of the color string.

Let us do one step further from the above six-quark picture to the properties of the $NN$ interaction.
The microscopic six-quark model predicts~\cite{Kusainov1991} that the
mixed-symmetry states $|s^4p^2[42]_xLST\rangle$ can be almost
confluent to the fully symmetric states $|s^6[6]\rangle$, so that, they can mix to
each other in the $S$-wave channels of $NN$ scattering. On the other hand, the
mixed-symmetry $6q$ components can overpass to the fully symmetric ones by the
$\sigma$ emission. If the emitted $\sigma$ meson has not very high energy and the
transition occurs in the field of the multi-quark core, the final $\sigma$ meson will attach to the
fully symmetric $6q$ core and this will lead to a significant energy
shift of the initial mixed-symmetry states\footnote{The similar attachment of the
$\sigma$ meson to the fully symmetric $|s^3[3]\rangle$ core leads to the well-known
mass shift of the Roper resonance which makes its Breit--Wigner mass (1440 MeV) ca. 500
MeV lower than the mass of the excited three-quark state $|sp^2[3]\rangle$.}.
This energy shift results
in a strong effective attraction in the respective $NN$ channels.

These findings form a base for the QCD-motivated dibaryon model of the $NN$ interaction which we discuss below. In the model, the basic intermediate-range attraction between nucleons is a consequence of the formation of a six-quark bag dressed by the strong scalar $\sigma$ field in $NN$ collisions. It may seem that the physical $\sigma$ meson (listed as $f_0(500)$ in the PDG tables~\cite{PDG}), which has a mass of about 500 MeV and a large width of the same order, can hardly play such a significant role in the $NN$ interaction. However, it was shown in, e.g., Refs.~\cite{Glozman,Volkov}, that the $\sigma$-meson mass and width can be strongly reduced, and thus it can become much more stable, due to the partial chiral symmetry restoration, which takes place in excited hadrons or dense baryon matter. We assume a similar mechanism to take place in the $6q$ states which satisfy both these conditions due to their compact size (at least for some of the known dibaryons) and inner $2\hbar \omega$ excitation. Thus, we have shown in Refs.~\cite{PRC2013,PRD2021} that the ABC effect in the reactions $pn \to d \pi\pi$ can be explained by the emission of the renormalized $\sigma$ meson with the mass of about $300$ MeV and width of about $100$ MeV from the $d^*(2380)$ dibaryon state. In the initial version of the dibaryon model for the $NN$ interaction~\cite{JPhys2001,IntJModPhys2002}, we formally dealt with the stable light scalar mesons with the mass of about $300$--$350$ MeV and zero width. We should note that the stable $\sigma$ meson as a pure phenomenological construction has been commonly used in the traditional meson-exchange models for the $NN$ interaction to account for the intermediate-range attraction. The intermediate dibaryon formation can at least partially substantiate the (relative) stability of the scalar mesons which arise not in the empty space between two nucleons, but in the field of $6q$ states. The direct inclusion of the $\sigma$ width in the model would strongly complicate the practical calculations and lead to arising of the complex potential, the imaginary part of which should be related to inelastic processes (mainly $2\pi$ production). In the present version of the model described below in Sec.~\ref{sec_recent}, we take into account the inelastic processes by introducing the dressed dibaryon width (which effectively includes the width of the $\sigma$ meson within the dibaryon).

\section{Quark degrees of freedom in the two-nucleon system}
\label{sec_quark}

The dibaryon concept for the $NN$ interaction, originally proposed in Ref.~\cite{PIYAF}, and the dressed bag model developed on its basis in Refs.~\cite{JPhys2001,IntJModPhys2002}, suggests the following picture of the interaction between nucleons.
 At relatively large distances ($r_{NN}>1 $~fm), nucleons interact by the traditional pion exchange. However, when nucleons approach each other at a distance of $r_{NN}\lesssim 1 $~fm, a compound dibaryon state is formed, which can be described as a six-quark bag, dressed by meson fields, where the most important one is a field of light scalar $\sigma$ mesons. As a result of multiple transitions of a two-nucleon system to the state of a dressed six-quark bag and vice versa, an effective interaction arises, which gives the main attraction between the nucleons at intermediate distances.

\subsection{Formal scheme including internal and external spaces}
\label{formal}

To describe such an interaction mechanism, it is convenient to use a two-channel formalism, which assumes that a system of two nucleons can be in two different states (channels): an external $NN$ channel and an internal dibaryon channel.
The total wavefunction of such a system consists of two components belonging to two different Hilbert spaces. Thus, it can be written as a two-component column:
\[\Psi \! \in \! {\cal H} = \left (\begin{array}{l} \Psi^{\rm ex}\in \! {\cal H}^{\rm ex}\\
\Psi^{\rm in}\in \! {\cal H}^{\rm in} \end{array} \right )\!.
\]
The two Hilbert spaces, ${\cal H}^{\rm ex}$ and ${\cal H}^{\rm in}$, have quite
different nature: $\Psi^{\rm ex}$ depends on
the relative coordinate (or momentum) of two nucleons and their spins, while
$\Psi^{\rm in}$ can depend on quark, gluon and meson variables of the internal state. The two independent
Hamiltonians are defined in each of these spaces:
 $h^{\rm ex}$ acts in ${\cal H}^{\rm ex}$  and
 $h^{\rm in}$ acts in ${\cal H}^{\rm in}$.

 The total Hamiltonian $h$ acting in the total Hilbert space
 ${\cal H}={\cal H}^{\rm ex} \oplus {\cal H}^{\rm in}$ can be written in a matrix form:
\begin{equation}
h = \left (\begin{array}{ll} h^{\rm ex} & h^{\rm ex,in}\\
 h^{\rm in,ex} & h^{\rm in} \end{array} \right )\!,
\label{h2}
\end{equation}
 where the transition operators $h^{\rm ex,in}= (h^{\rm in,ex})^\dagger$ determine the coupling between external and internal channels. Note that if operators $h^{\rm ex}$ and
 $h^{\rm in}$ are self-adjoint and $h^{\rm ex,in}$ is bounded, then the Hamiltonian
 $h$ is the self-adjoint operator in $\cal H$.

The external Hamiltonian
\[h^{\rm ex} = t + v^{\rm ex} \]
includes the kinetic energy $t$ of the $NN$ relative motion and some peripheral part of the interaction $v^{\rm ex}$, i.e., the peripheral part of the meson-exchange potential and the Coulomb interaction in the case of two protons.

The total wavefunction $\Psi$ satisfies the two-component Schr\"odinger equation
  \begin{equation}    h\Psi= E\Psi.
   \label{Schr_2ch}
 \end{equation}
By excluding the internal component, one
 obtains an effective Schr\"odinger equation for the external channel only:
 \begin{equation}  h^{\rm eff}(E)\Psi^{\rm ex} =E\Psi^{\rm ex}
 \label{Sch2}
 \end{equation}
with an effective ``pseudo-Hamiltonian''
\begin{equation}
h^{\rm eff}(E) = h^{\rm ex}+h^{\rm ex,in}\,g^{\rm in}(E)\,h^{\rm in,ex} = t +v^{\rm ex} + w(E),
\label{heff2}
\end{equation}
which depends on energy\footnote{From the
mathematical point of view, an operator depending on the spectral
parameter is not an operator at all, because its domain depends on
this spectral parameter. Therefore, strictly speaking, such an object should not be called the Hamiltonian. However, physicists ignore this fact and use energy-dependent interactions very widely.} $E$ due to the
resolvent of the internal Hamiltonian $g^{\rm in}(E)=(E-h^{\rm in})^{-1}$.

 Having found the solution $\Psi^{\rm ex}$ of the effective equation~(\ref{Sch2}), one
 can uniquely restore the excluded internal state:
 \begin{equation}
 \Psi^{\rm in} =g^{\rm in}(E)h^{\rm in,ex}\Psi^{\rm ex}.
 \label{psiin}
 \end{equation}

To determine the components of the total Hamiltonian in the above formal scheme, it is necessary to use some microscopic theory, which, in principle, is able to describe both the external and internal channels and, most importantly, the transitions between them. A six-quark model was used for these purposes in Refs.~\cite{JPhys2001,IntJModPhys2002}. We briefly outline below the main assumptions and the resulting form of the dibaryon model following from the microscopic six-quark treatment of the $NN$ system.

\subsection{Six-quark structure of the two-nucleon system: symmetry aspects}
\label{6qNN}

Within the microscopic six-quark description, the RGM ansatz
can be used for the $NN$-channel wavefunction:
 \begin{equation}
  \Psi_{NN}^{\rm RGM}(123456)=
  {\cal A}\{\psi_N(123)\psi_N(456)\chi_{\scriptscriptstyle NN}({\bf r})\},
\label{RGM}
\end{equation}
where ${\bf r}=\frac{1}{3}({\bf r_1}\!+\!{\bf r_2}\!+\!{\bf r_3}\!-\!{\bf r_4}\!-\!{\bf r_5}\!-\!{\bf r_6})$ is the
distance between the nucleon clusters, $\psi_N(i,j,k)$ is the quark wavefunction of the nucleon:
 \begin{equation}
\!\!\psi_N(\!123)\!=\!\varphi_N(\!{\bm \rho}_1,{\bm\xi}_1\!)\,|[1^3]_CS_{3q},
\!([21]_{CS}\!)T_{3q}\!:\![1^3]_{CST}\!\rangle,
 \label{n}
 \end{equation}
with ${\bm \rho}_1={\bf r}_1-{\bf r}_2$,
${\bm \xi}_1=\frac{1}{2}({\bf r}_1+{\bf r}_2)-{\bf r}_3$,
$S_{3q}=1/2$, $T_{3q}=1/2$,
and ${\cal A}$ is the antisymmetrizer consisting of permutations of all six quarks.

Then the wavefunction in the external channel corresponds to the
renormalized RGM relative-motion function~\cite{Saito1969}:
 \begin{equation}
  \Psi^{\rm ex}({\bf r}) \to {\cal N}^{-1/2}\chi_{\scriptscriptstyle NN}({\bf r}),
\label{psinn}
\end{equation}
where ${\cal N}$ is the so-called overlap kernel:
 \begin{equation}
  {\cal N}({\bf r',r})
  = \langle\psi_N\psi_N|{\cal A}\delta({\bf r}^\prime-{\bf r})
  |\psi_N\psi_N\rangle.
\label{overl}
\end{equation}

The external and transition Hamiltonians correspond to the following RGM expressions:
 \begin{eqnarray}
&h^{\rm ex}\to h^{\rm ex}_{\scriptscriptstyle RGM}({\bf r}^\prime,{\bf r})=
\langle\psi_N\psi_N|{\cal A}h_{6q}^{\rm ex}|\psi_N\psi_N\rangle,\nonumber\\
&h^{\rm ex,in}\to h^{\rm ex,in}_{\scriptscriptstyle RGM}
({\bf r}^\prime;{\bf r},\!\{{\bm\rho}{\bm\xi}\})=\langle\psi_N\psi_N|{\cal A}h_{6q}^{\rm ex,in},
\label{hrgm}
\end{eqnarray}
which include some microscopic $6q$ Hamiltonian. Here, for brevity, the set of inner
coordinates of the six-quark system ${\bf r},{\bm\rho}_1,{\bm\xi}_1,{\bm\rho}_2,{\bm\xi}_2$
is denoted by ${\bf r},\!\{{\bm\rho}{\bm\xi}\}$.

Next, we consider the possible symmetry of the $6q$ wavefunctions in the framework of the translationally invariant shell model (TISM) including all $6q$ configurations with 0$\hbar\omega$, 1$\hbar\omega$ and
2$\hbar\omega$ excitations.
Let us consider the possible $6q$ spatial symmetries of the
external $NN$ channel, e.g., in the case of an $S$ partial wave. If one
assumes the symmetry of the nucleon wavefunction as $[f_X]=[3]_X$, then
the allowed $6q$ symmetries in even partial waves should be $[6]_X$ and
$[42]_X$. These two
components should be associated with unexcited $|s^6[6]_XL=0\rangle$
and excited $|s^4p^2[42]_XL=0,2\rangle$ configurations, respectively. It
was shown in Refs.~\cite{Kusainov1991,Obukhovsky1996} that
the components of the first type should be identified with bag-like
configurations, while the second-type components can be naturally identified
with the proper $NN$ configurations.

The quark wavefunction of the nucleon is a quark shell-model configuration $s^3[3]_X$
symmetric in the $X$ (coordinate) space, and in the CST (color, spin, isospin) space.
It is described by a specific set of Young schemes satisfying the Pauli principle:
\begin{equation}
\begin{split}
&|N\rangle=|s^3[3]_X\rangle|N(CST)\rangle,\\
|N(CST)\rangle\!=&|[1^3]_{CST}\{[21]_{CS}([1^3]_C\!\circ\![21]_S)\!\circ\![21]_T\!\}\rangle
\end{split}
\label{q1}
\end{equation}

The $6q$ wavefunction of the $NN$ system composed from the free-nucleon states
(\ref{q1}) can be expanded (see the formalism in Ref.~\cite{Harvey1979}) in the
quark shell-model configuration series which includes both fully symmetric
$s^6[6]_X$ and mixed-symmetry $s^5p[51]_X$, $s^4p^2[42]_X$, $s^3p^3[3^2]_X$
states. Here the $s^6$ and $s^4p^2$ configurations correspond to the even $NN$ partial waves, while
$s^5p$ and $s^3p^3$ ones --- to the odd $NN$ partial waves. It is important that the restrictions
imposed by the Pauli principle in the mixed-symmetry states are not as stringent as
in the case of the fully symmetric ones $[6]_X$, but still remain quite important
for the ``almost symmetric'' Young scheme $[51]_X$.

The following basic sets of states satisfy the Pauli principle for the $NN$
channels~\cite{Obukhovsky1979,Kusainov1991} (taking into account the
requirement of color neutrality, $[1^3]_C \times [1^3]_C \to [2^3]_C$):

1. For even orbital momenta $L$ (i.e., $ST=10,\,01$)
\begin{equation}
\begin{split}
\Psi_0&=|s^6[6]_XL\!=\!0;[1^6]_{CST}\\
&\times\!\{[2^3]_{CS}([2^3]_C\!\circ\![42]_S)\!\circ\![3^2]_T\}\rangle,\,\,ST=10,
\end{split}
\label{q2}
\end{equation}
\begin{equation}
\begin{split}
\Psi^\prime_0&=|s^6[6]_XL\!=\!0;[1^6]_{CST}\\
&\times\!\{[2^21^2]_{CS}([2^3]_C\!\circ\![3^2]_S)\!\circ\![42]_T\}\rangle,\,\, ST=01,
\end{split}
\label{q2prime}
\end{equation}
and
\begin{equation}
\begin{split}
\Psi_2^{(i)}&=|s^4p^2[42]_XL\!=\!0,2;[2^21^2]_{CST}\\
&\times\!\{[f_2]_{CS}([2^3]_C\!\circ\![42]_S)\!\circ\![3^2]_T\}\rangle,\,\, ST=10,
\end{split}
\label{q3}
\end{equation}
where $i=1,2,...,5$ is the number in the set of color-spin ($CS$) Young schemes:
\begin{equation}
[f_2]_{CS}(\!=\![2^3]_C\!\circ\![42]_S)\!=[42],\,[321],[2^3],[31^3],[21^4]
\label{q3a}
\end{equation}
(here $[2^3]_C\!\circ\![42]_S$ is the inner product of the color and spin Young
schemes),
\begin{eqnarray}
\begin{split}
{\Psi^\prime}_2^{(i)}&=|s^4p^2[42]_XL\!=\!0,2;[2^21^2]_{CST}\\
&\times\!\{[f^\prime_2]_{CS}([2^3]_C\!\circ\![3^2]_S)\!\circ\![42]_T\}\rangle,\,\, ST\!=\!01,
\end{split}
\label{q3prime}
\end{eqnarray}
with $i=1,...,4$ being the number in the set
\begin{equation}
[f^\prime_2]_{CS}(\!=\![2^3]_C\!\circ\![3^2]_S)\!=[3^2],\,[41^2],\,[2^21^2],\,[1^6].
\label{q3b}
\end{equation}

2. For odd orbital momenta $L$ (i.e., $ST=11,\,00$)
\begin{eqnarray}
\begin{split}
\Psi_1\!&=|s^5p[51]_XL\!=1;[21^4]_{CST}\\
&\times\!\{[\tilde f_1]_{CS}([2^3]_C\!\circ\![42]_S)\!\circ\![42]_T\}\rangle,\,\,ST\!=\!11
\end{split}
\label{q4}
\end{eqnarray}
(here only the values $[\tilde f_1]_{CS}=[321],[2^3],[21^4]$ from the set (\ref{q3a}) are allowed,
while $[42]$ and $[31^3]$ are forbidden),
\begin{eqnarray}
\begin{split}
\Psi^\prime_1\!&=|s^5p[51]_XL\!=1;[21^4]_{CST}\\
&\times\!\{[2^21^2]_{CS}([2^3]_C\!\circ\![3^2]_S)\!\circ\![3^2]_T\}\rangle,\,\,ST\!=\!00,
\end{split}
\label{q4prime}
\end{eqnarray}
and
\begin{eqnarray}
\begin{split}
\Psi_3^{(i)}&=|s^3p^3[3^2]_XL=1,3;[2^3]_{CST}\\
&\times\!\{[f_3]_{CS}([2^3]_C\!\circ\![42]_S)\!\circ\![42]_T\}\rangle,\,\,ST\!=\!11
\end{split}
\label{q5}
\end{eqnarray}
with $i=1,2,...,5$ being the number in the set (\ref{q3a}),
\begin{eqnarray}
\begin{split}
{\Psi_3^\prime}^{(i)}&=|s^3p^3[3^2]_XL=1,3;[2^3]_{CST}\\
&\times\!\{[f^\prime_3]_{CS}([2^3]_C\!\circ\![3^2]_S)\!\circ\![3^2]_T\}\rangle,\,\,ST\!=\!00
\end{split}
\label{q5prime}
\end{eqnarray}
with $i=1,...,4$ being the number in the set (\ref{q3b}).

One can see that the Pauli exclusion principle does not forbid the unexcited configurations
$s^6[6]_X$ and $s^5p[51]_X$ in the channels with positive and
negative parity, respectively, but severely limits the set of allowable Young schemes in
the $CS$ subspace, reducing the allowed basis to only a single $CS$ state $[2^3]_{CS}$
($[2^21^2]_{CS}$) in even triplet (singlet) partial waves and strongly restricting the basis
in odd partial waves. At the same time, the excited configurations $s^4p^2[42]_X$ and
$s^3p^3[3^2]_X$ satisfy the Pauli principle for any value of the $CS$ Young
scheme from the Clebsch--Gordan series (\ref{q3a}) and (\ref{q3b}) for the inner product of
color and spin Young schemes in the triplet ($S=\,$1) and singlet ($S=\,$0) channels.
So, in a rough approximation, one can evaluate the short-range $NN$ interaction
by considering the configurations dominating in the overlap region of two
nucleons.

In quark models that use the QCD-induced interaction~\cite{DeRujula1975}, viz.,
the confinement potential $\sim \lambda_i\lambda_j$ and the spin-dependent
color-magnetic interaction $\sim \lambda_i\lambda_j \sigma_i\sigma_j$, the state
with the most symmetric Young scheme $[42]_{CS}$ from the series (\ref{q3a}) is
marked out in energy. Note that the energy splitting between the states with the color-spin symmetry
$[42]_{CS}$ and $[2^3]_{CS}$ is of the order of magnitude of the $N-\Delta$
splitting (i.e., the splitting between the hadronic states with the color-spin symmetry
$[21]_{CS}$ and $[1^3]_{CS}$).
It is important that, from the whole series (\ref{q3a}), only the first term $[42]_{CS}$
corresponds to the state, in which the color-magnetic interaction term leads to the
$NN$ attraction in the overlap region~\cite{Obukhovsky1979,Kusainov1991}.
In the singlet channel, the state with the
most symmetric Young scheme $[3^2]_{CS}$ from the series (\ref{q3b}) plays the same role.
Consequently, the dominance of the configurations $s^4p^2[42]_X $ and $s^3p^3[3^2]_X$
over the more symmetric ones $s^6[6]_X$ and $s^5p[51]_X$ in the overlap area can
lead to the $NN$ attraction instead of the strong short-range repulsion in the
traditional approaches.

The numerical calculations~\cite{Kusainov1991} of $NN$ elastic scattering within
the RGM framework confirmed the above
conclusions made from the symmetry considerations. In these calculations, the
authors used the QCD-induced interaction and took into account the exchange of
the Goldstone boson \{$\sigma, \bm{\pi}$\} between quarks.

The six-quark RGM wavefunction of the $NN$ system $\Psi_{NN}={\cal
A}\{\chi_{\scriptscriptstyle NN}(r;E)N(123)N(456)\}$ corresponding to the
realistic description of the scattering phase shifts in the $^3S_1$ wave in a wide energy range
$0< E \lesssim 1$~GeV was projected onto the shell-model $6q$
configurations $s^6[6]_X $ and $s^4p^2[42]_X$ by using the TISM methods. As a result, the following important representation was obtained for the microscopic wavefunction
$\Psi_{NN}$~\cite{Kusainov1991}:
\begin{eqnarray}
\begin{split}
&\Psi_{NN}(^3\!S_1;E)=c_0(E)\Psi_0+\Psi^Q_{NN}(^3\!S_1;E),\\
&\Psi^Q_{\!NN}(^3\!S_1;E)=\sum_{i=1}^5c_2^{(i)}\!(E)\Psi_2^{(i)}\!+\!
{\cal A}\{\chi^{ass}_{\scriptscriptstyle NN}(r;E)NN\},
\end{split}
\label{q7}
\end{eqnarray}
where the antisymmetrizer $\cal{A}$ was omitted before $6q$ configurations $\Psi_n$, since the
basic states (\ref{q2})--(\ref{q5prime}) are antisymmetric by definition:
\begin{equation}
{\cal A}\Psi_n=\Psi_n,\quad  {\cal A}^2={\cal A}.
 \label{apsi}
\end{equation}
Note that the same expansion can also be written for
the singlet $S$-wave $NN$ channel:
\begin{eqnarray}
\begin{split}
&\Psi^\prime_{NN}(^1\!S_0;E)=c^\prime_0(E)\Psi^\prime_0+
{\Psi^\prime}^Q_{NN}(^1\!S_0;E),\\
&{\Psi^\prime}^Q_{\!\!NN}(^1\!S_0;\!E)\!=\!
\sum_{i=1}^4\!c_2^{\prime(i)}\!(\!E)\Psi_2^{\prime(i)}\!+\!
{\cal A}{\{\chi^\prime}^{ass}_{\!\!\scriptscriptstyle NN}(r;E)NN\}.
\end{split}
\label{q7prime}
\end{eqnarray}
In both cases, the first term proportional to $\Psi_0$($\Psi^\prime_0$) includes
a coherent superposition of $NN$, $\Delta$--$\Delta$ and $CC$ states
(see, e.g., the first column of Tab.~\ref{TabA} in Appendix~\ref{formfactor}) with the
large weight just for the $CC$ component (states with the hidden color). So that,
this term likely corresponds to a $6q$ bag-like component.

The second term $\Psi^Q_{NN}$ [${\Psi^\prime}^Q_{NN}$] includes a coherent superposition
of five [four] components corresponding to the mixed-symmetry configurations
 $s^4p^2[42]_X$ [$s^3p^3[3^2]_X$] with all $CS$ Young schemes from the series
 (\ref{q3a}) [(\ref{q3b})].
This term has been demonstrated~\cite{Kusainov1991} to correspond to a
state vector where the cluster $NN$ component (i.e., widely spaced and non-symmetrized
product of the nucleonic wave functions) has the maximal weight, while the remain
components interfere destructively and, as a result, can only be considered as small
corrections to the basic $NN$ component. At the same time, the asymptotic part
${\cal A}\{\chi^{ass}_{\scriptscriptstyle NN}(r;E)N(123)N(456)\}$ of the cluster
component $\Psi^Q_{NN}$ is orthogonal to configurations $\Psi_n$ and has only a minor
effect on the short-range wavefunction~\cite{Kusainov1991}.

Thus, according to Ref.~\cite{Kusainov1991}, the quark-model wavefunction of $NN$
scattering in the $^3S_1$ partial wave consists of two qualitatively different components:
the shell-model state $s^6[6]_X$ symmetric in the coordinate space, like a $6q$ bag
composed mainly from $CC$ states and corresponding to the internal dibaryon channel,
and the $NN$ cluster-like state $\Psi^Q_{NN}$ corresponding to $2\hbar\omega$-excited
relative motion of two nucleons at short distances (i.e., the external channel).
By analogy, we may assume that the quark-model wavefunction of $NN$ scattering in
the $^1S_0$ partial wave also consists of two different components: the shell-model
bag-like state $s^6[6]_X$ and the $NN$ cluster-like state ${\Psi^\prime}^Q_{NN}$.
Therefore, the transition from the external $NN$ component (mainly having the mixed
symmetry) to the internal $6q$ bag components must be accompanied by a transition
of two $p$-shell quarks to the $s$ shell with an emission of two tightly correlated
pions. 

The projection of the cluster component onto the $NN$ channel in the overlap
region $r \lesssim 2b$ (where $b$ is the radius of nucleon ``quark core'') at each
fixed value of energy $E$ takes the form:
\begin{equation}
\sqrt{\frac{6!}{3!3!2!}}\langle NN|\sum_{i=1}^5\!c_2^{(i)}|\Psi_2^{(i)}\rangle\!=\!
{\cal N}_0(1\!-\!\frac{r^2}{b^2})\exp(-\frac{3r^2}{4b^2})
 \label{q9}
\end{equation}
(we should use the primed terms $c_2^{\prime(i)} \Psi_2^{\prime(i)}$ in the case of the
$^1\!S_0$ wave). This relative-motion wavefunction has a radial node localized at the
distance $b$. According to Ref.~\cite{Kusainov1991},
both components have approximately equal probabilities $\sum_{i =
1}^5|c_2^{(i)}(E)|^2 \approx |c_0(E)|^2$ for any value of $E$ in the interval
$0<E \lesssim 1$~GeV. Consequently, the node of the $NN$ cluster part of the
wavefunction (\ref{q9}) is almost independent on energy in this range.
This means that only the normalization factor ${\cal N}_0$ in the r.h.s. of
Eq.~(\ref{q9}) is really dependent on energy up to $\approx 1$ GeV.

\subsection{The nodal structure of $NN$ wavefunctions}
\label{node}

The stationary node at the distances $r\approx b$ plays the same role in $NN$
elastic scattering as the repulsive core in the traditional potential models for the $NN$
interaction. In particular, the stationary node of the $NN$ wavefunction
completely explains the constant negative slope of the phase shifts in the
$^3S_1$ ($^1S_0$) partial wave up to energies $E\approx 1$~GeV.

As for the $^3\!D_J$ ($^1\!D_2$) partial waves, they correspond to the configuration
$s^4p^2[42]_X(L\!=\!2)$ in the quark-model description. So, in the $^3\!D_J$ ($^1\!D_2$)
channels, instead of Eqs.~(\ref{q7})--(\ref{q7prime}), one gets a similar expansion:
\begin{eqnarray}
&\Psi_{NN}(^3\!D_J;E)=\Psi^Q_{NN}(^3\!D_J;E)\!=\!
\nonumber\\
&\sum_{i=1}^5\!d_{2J}^{(i)}(E)\Psi_{2DJ}^{(i)}\!+\!
{\cal A}\{\chi^{ass}_{\scriptscriptstyle \!N\!ND\!J}(r;E)Y_2(\hat r)NN\}_{\!J}
\label{q12}
\end{eqnarray}
for the triplet $D$ waves and for the singlet $D$ wave as well (with primed basis functions and
primed coefficients).

Despite the fact that the expansions (\ref{q7}) and (\ref{q12}) look to be very similar,
they nevertheless correspond to a different behavior of $NN$ scattering phase shifts.
This can be seen by projecting the cluster component $\Psi^Q_{NN}(^3\!D_J;E)$
onto the $NN$ channel. Instead of the nodal function (\ref{q9}), one gets a nodeless radial function
\begin{eqnarray}
\sqrt{\frac{6!}{3!3!2!}}\langle NN|\!\sum_{i=1}^5\!d_2^{(i)}|\Psi_{2D}^{(i)}\rangle\!=\!
{\cal N}_2\frac{r^2}{b^2}Y_2(\hat r)\exp(-\frac{3r^2}{4b^2})
\label{q13}
\end{eqnarray}
Therefore, in contrast to the $^3\!S_1$ ($^1\!S_0$) wave, the hard-core effects should not be
manifested in the energy dependence of the $^3\!D_J$ ($^1\!D_2$) phase shift. However, the
spin-orbit interaction and tensor coupling in the triplet channel could modify the energy
dependence of the $^3\!D_J$ phase shift as compared to our qualitative quark-model consideration with account of the Pauli exclusion principle.

The situation with the triplet odd partial waves $^3\!P_J$ and $^3\!F_J$ looks a
little bit more complicated than for even partial waves. In fact, the possible
Young schemes for the orbital part of the $6q$ wavefunction in $P$ waves
are $s^5p[51]_X$ and $s^3p^3[33]_X$ with respective $CS$ parts shown in
Eqs.~(\ref{q4})--(\ref{q5prime}). So, we can assume that the wavefunction with the symmetry
 $s^5p[51]_X$, being the intermediate between the fully symmetric bag-like component
 $s^6[6]_X$ and the highly clusterized state $s^4p^2[42]_X$, should manifest some cluster-like
 properties, i.e., it can dissociate into the respective $NN$ channel.

On the other hand, in $P$ waves one should take into consideration the spin-orbit
splitting even for the $6q$ wavefunction. Taking into account the spin-orbit
splitting, one notes that $p_{\tralf}$ quark orbit lies lower than $p_{\half}$
orbit. So, the $6q$ configuration $s^5p[51]_X$ includes just the $p_{\tralf}$
single-quark orbit, i.e., it corresponds to the $s^5p_{\tralf}[51]_X$ state,
quite similarly to the nuclear physics case of the $^5$Li and $^5$He ground
states (with the nuclear shell-model configuration $s^4p_{\tralf}[41]_X$).
In turn, the possible total angular momenta for the configuration
$s^5p_{\tralf}[51]_X$ are $J=2$ and $J=1$, but not $J=0$. In such a case, in the triplet
$NN$ channels $^3\!P_2$--$^3\!F_2$ and $^3P_1$, one has a superposition of two
$6q$ components: $s^5p_{\tralf}[51]_X +s^3p^3[33]_X$, while in the $^3P_0$
channel one has the nodal $s^3p^3[33]_X$ component only. Due to the presence of a
radial node in the $NN$ scattering wavefunction, this component corresponds to the
strongly enhanced kinetic energy and thus induces
some additional repulsion in the $NN$ system.

Although, in general, the formalism for the odd partial waves $^3\!P_J$--${}^3\!F_J$ and $^1\!P_1$--${}^1\!F_3$ is more complicated, we use here the basis vectors from the set (\ref{q4})--(\ref{q5prime}), in the same way as we used the states (\ref{q2})--(\ref{q3prime}) for the even partial waves
$^3\!S_1$--${}^3\!D_J$ and $^1\!S_0$--${}^1\!D_2$. First, we consider the basis of the state with the total orbital momentum $L = 1$. The state with the Young scheme $[42]_{CS}$ apparently plays
here the same role as in the channels $^3S_1$--${}^3D_J$. This can be seen from
the comparison of the basis vectors (\ref{q4}), (\ref{q5}) and (\ref{q2}), (\ref{q3}). Therefore,
it is also possible here to expand the quark wavefunction into an $NN$ cluster part and a bag-like
part with the more symmetric wavefunction in the coordinate space (its CST content is strongly limited by the Pauli principle):
\begin{eqnarray}
\begin{split}
&\Psi_{N\!N}(^3\!P_J;E)=\sum_{i\!=\!1}^3\tilde p^{(i)}_{1J}(E)\Psi^{(i)}_{1J}+
\Psi^Q_{\!N\!N}(^3\!P_J;E),\\
&\Psi^Q_{N\!N}(^3\!P_J;E)\!=\!\sum_{i=1}^5\!p_{3J}^{(i)}(E)\Psi_{3J}^{(i)}\!+\!
{\cal A}\{\chi^{ass}_{\scriptscriptstyle \!N\!N\!P\!J}(r;\!E)Y_1(\hat r)NN\}_{\!J}.
\end{split}
\label{q14}
\end{eqnarray}
Here $\tilde p^{(i)}_{1J}(E)$ and $p^{(i)}_{3J}(E)$ are the expansion coefficients of the $NN$-scattering quark wavefunction (e.g., the solution of the MRG equation) with a given value of the total angular momentum $J=L+S$ (here $S=1$ and the obvious algebra of addition of momenta is omitted). In the overlap area $r\lesssim 2b$, the projection of the cluster component of the function (\ref{q14}) onto the $NN$ channel, calculated by the TISM methods, must have a node at a distance $r = b$, regardless of the specific value of $J$, if there is no spin-orbit interaction.
\begin{equation}
\sqrt{\!\frac{6!}{3!3!2!}}\langle NN|\!\sum_{i=1}^5\!p_{3J}^{(i)}|\Psi_{3J}^{(i)}\rangle\!=\!
{\cal N}_3(\!1\!-\!\frac{r^2}{b^2})\frac{r}{b}Y_1(\hat r)\exp(-\frac{3r^2}{4b^2})
\label{q15}
\end{equation}
In reality, the spin-orbit interaction is present, and, as a result, the phase shifts in $^3\!P_J$
channels have significant splitting in $J$. Moreover, $^3\!P_0$ and $^3\!P_1$ phase shifts have a
constant negative slope, which indicates the presence of a stable node in the cluster component of
the quark wavefunction and the predominance of an attractive force at small distances. However, the
behavior of the $^3\!P_2$ phase shift is different, which is possibly explained by the essential role
of tensor mixing $^3\!P_2$--${}^3\!F_2$.

On the other hand, if to rewrite the shell-model $6q$ wavefunctions
$s^5p[51]_X$ and $s^3p^3[33]_X$ within the framework of the Nijmegen--ITEP $4q-2q$ model~\cite{Nijm,ITEP} (see Sec.~\ref{ITEP}),
then one gets a strong spin-orbit attraction in the
component $|s^5p[51]_X, J=2\rangle$ and much weaker --- in the component
$|s^5p[51]_X, J=1\rangle$. So, the weight of the nodeless component $|s^5p[51]_X,
J=2\rangle$ should be much higher than the weight of the $|s^5p[51]_X, J=1\rangle$ component,
as compared with the contribution of the second (nodal) component $|s^3p^3[33]_X, J\rangle$.

As a result of this qualitative consideration, one can conclude that the $NN$ radial
wavefunction in the triplet $^3\!P_2$ channel should be predominantly nodeless with the
main component $|s^5p[51]_X, J=2\rangle$. At the same time, for the $^3\!P_1$ channel, the situation is
opposite, i.e., the $|s^3p^3[33]_X, J=1\rangle$ component should dominate. In the
$^3\!P_0$ channel, the mixed-symmetry configuration $|s^3p^3[33]_X,
J=0\rangle$ leads to the nodal radial wavefunctions. The empirical behaviour of the triplet $P$-wave phase
shifts correspond exactly to such a behaviour of quark wavefunctions, which follows from the
microscopic consideration.

The situation in the triplet $^3\!F_J$ channels resembles one in the $^3\!P_J$
channels: the $p$-shell states with the mixed-symmetry configuration $|s^3p^3[33]_X,
J\rangle$, $J=2,3,4$, correspond to the nodal radial wavefunction, but in this case
there should be an admixture of the $f$-shell states with an almost symmetric
configuration $|s^5\!f[51]_X, J\rangle$ having the $CS$ structure considerably
restricted by the Pauli principle.

Hence, all the triplet odd partial waves $^3\!P_J$ and $^3\!F_J$ can
be described by the dibaryon model for the $NN$ interaction. The situation in
the singlet odd channels $^1\!P_1$ and $^1\!F_3$ is simpler, since the
spin-orbit splitting is absent here and there is only one bag-like state
(\ref{q4prime}) in the $P$ wave, while the expansion similar to Eq.~(\ref{q14}) here
has the form:
\begin{eqnarray}
\begin{split}
\Psi_{NN}(^1\!P_1;E)&=p^\prime_1(E)\Psi^{\prime}_1+\Psi^Q_{NN}(^1\!P_1;E),\\
&\Psi^Q_{NN}(^1\!P_1;E)=\\
\sum_{i=1}^4 p_3^{\prime(i)}(E)\Psi_3^{\prime(i)}&\!+\!
{\cal A}\{\!\chi^{ass}_{\scriptscriptstyle N\!N\!P\!1}(r;\!E)Y_1(\hat r)NN\}_{\!J\!=\!1}.
\end{split}
\label{q16}
\end{eqnarray}
An analogous expansion is valid for $\Psi_{NN}(^1\!F_3;E)$ with the substitution
$p^\prime_1(E) \to f^\prime_1(E)$, $\Psi^{\prime}_1 \to \Psi^{\prime\prime}_1=
|s^5\!f[51]_XL\!=3\rangle$, \ldots, etc.

Thus, we have shown that just the mixed-symmetry states
with the six-quark structure $|s^4p^2[42]_xLST\rangle$ dominate over the fully space-symmetric configuration $|s^6[6]\rangle$ due to a much higher statistical weight and
also specific features of the quark-quark interaction ($v_{qq}\sim
\vec{\lambda}_i\vec{\lambda}_j\vec{\sigma}_i\vec{\sigma}_j $). When treating the
$S$-wave $NN$ interaction, this property leads to the presence of a stationary node in the $NN$ radial wavefunctions in a broad energy range from zero to 1 GeV~\cite{Kusainov1991}.

The same property is valid also for the $P$-wave states of the $6q$ system where the
$|s^3p^3[33]_xLST\rangle$ configuration dominates over the
$|s^5p[51]_xLST\rangle$ one. So, the $NN$ scattering radial wavefunctions in such $P$ waves must display a similar
nodal behavior.

\section{Effective Hamiltonian of the dibaryon model}
\label{sec_effham}
Let us now turn to the formalism of the dibaryon model. After excluding the internal channel from the total Hamiltonian (\ref{h2}) acting in the two-component Hilbert space, one comes to an effective $NN$ Hamiltonian (\ref{heff2}):
\begin{equation}
h^{\rm eff}(E) = h^{\rm ex} + w(E),
\end{equation}
where $w(E)=h^{\rm ex,in}\,g^{\rm in}(E)\,h^{\rm in,ex}$ is the effective energy-dependent interaction due to the coupling with the internal channel. We remind that the external Hamiltonian $h^{\rm ex}$ includes, in addition
to the kinetic energy $t$, the peripheral
meson-exchange interaction.
Below we show how the explicit form of the effective interaction $w(E)$ can be obtained from the microscopic quark consideration in the pole approximation.

\subsection{The pole approximation for the resolvent of the internal Hamiltonian}
\label{pole}

A complete description of the internal channel as a system of six interacting quarks
surrounded by a meson field is a complicated problem. However, to determine the
effective potential $w(E)$ in the external $NN$ channel, it is sufficient to take into account only one or few lowest states of the $6q$ bag, and this leads to a simple pole approximation for the resolvent of the internal channel:
\begin{equation}
g^{\rm in}(E)=\sum_{\alpha}\int\frac{|\alpha,{\bf k}\rangle
\langle \alpha,{\bf k}|d^3k}{E-E_{\rm in}(\alpha ,{\bf k})},
\label{resn}
\end{equation}
where $|\alpha\rangle$ is the $6q$ part of the wavefunction for the
dibaryon states and the plane waves $|{\bf k}\rangle$ describe
the $\sigma$-meson states. The total energy
$E_{\rm in}(\alpha,{\bf k})$ of the dressed bag is
 \begin{equation}
 E_{\alpha }({\bf k}) = m_{\alpha}+\varepsilon_{\sigma}(k),
 \label{eak}
\end{equation}
where
  \begin{equation}
 \varepsilon_{\sigma}(k) = k^2/2m_{\alpha}+\omega_{\sigma}(k) \simeq
 m_{\sigma}+k^2/2\bar{m}_{\sigma},
 \label{epssig}
\end{equation}
 $\omega_{\sigma}(k)=\sqrt{m_{\sigma}^2+k^2}$ is relativistic energy
 of the $\sigma$ meson, $\bar{m}_{\sigma}=
 m_{\sigma}m_{\alpha}/(m_{\sigma}+m_{\alpha})$ is the reduced mass of the dressed bag and $m_{\sigma}$ and $m_{\alpha}$ are the masses of the $\sigma$ meson
 and the bare $6q$ bag, respectively.

\begin{figure}[h]
\begin{center}
 \epsfig{file=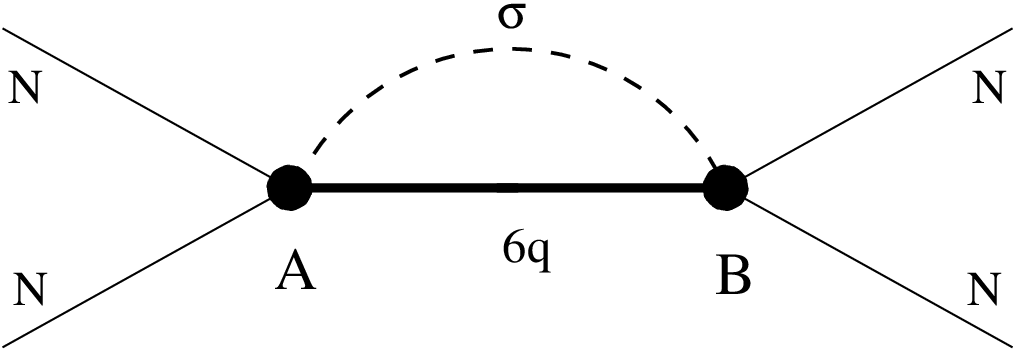,width=0.8\columnwidth}
\end{center}
\caption{Effective $NN$ interaction induced by the formation of an
intermediate $6q$ bag dressed by meson fields.}
\label{f1}
\end{figure}

Using the pole approximation (\ref{resn}) for  $g^{\rm in}$, one can present the effective interaction $w(E)$ as a sum of factorized terms:
 \begin{equation}
w(E)=\sum_{\alpha}\int\frac{h^{\rm ex,in}|\alpha,{\bf k}\rangle
\langle \alpha,{\bf k}|h^{\rm in,ex}\,d^3k}{E-E_{\rm in}(\alpha ,{\bf k})}.
\label{we}
\end{equation}
 Such an effective interaction $w(E)$ resulted from the coupling of
the external $NN$ channel to the dressed $6q$ bag is
illustrated by the graph in Fig.~\ref{f1}.

Since we have adopted the pole approximation (\ref{resn}) for the resolvent of the
internal channel $g^{\rm in}$, the derivation of the effective interaction $w$ in
the external channel does not require the knowledge of the full internal Hamiltonian
$h^{\rm in}$ of the dressed bag, as well as the full transition operator $h^{\rm
ex,in}$. As follows from Eq.~(\ref{we}), it is necessary to determine the result of the
action of the transition operator only on those states of the dressed bag
$|\alpha,\bf k\rangle$ which we included in the resolvent $g^{\rm in}$.

If the RGM ansatz~(\ref{RGM}) is used to describe the external channel, then the action
of the transition operator~(\ref{hrgm}) on the state $|\alpha,\bf k\rangle$ can be formally written as
\begin{equation}
h^{\rm ex,in}|\alpha,{\bf k}\rangle=
\langle\{\psi_N\psi_N\}_{ST}|\hat O_{\sigma}({\bf k},E)
|\alpha,{\bf k}\rangle,
\label{tr1}
\end{equation}
where $\hat O_{\sigma}({\bf k},E)$ is an annihilation operator for
the $\sigma$ meson with the momentum ${\bf k}$ and the indices $ST$ are
the spin-isospin quantum numbers of the $NN$ state [with the nucleon $3q$ wave
function given by Eq.~(\ref{n})].

The formal expression (\ref{tr1}) already allows us to define the general structure of
the transition vertex $B$ in Fig.~\ref{f1} without detailing the form of the operator $\hat
O_{\sigma}({\bf k},E)$. After the partial-wave decomposition of the r.h.s. of
Eq.~(\ref{tr1}), one obtains (at a fixed orbital momentum of the $\sigma$ meson
$L_{\sigma}$):
\begin{equation}
h^{\rm ex,in} |\alpha\{S_{\alpha},L_{\sigma}\}_{JM},k\rangle=
\sum_L  B_{L_{\sigma}LS}^J(k,E)\!|Z^{JM}_{LS}\rangle,
 \label{tr2}
 \end{equation}
where the sum over $L$ includes all the admissible values of $NN$
orbital momenta compatible with the fixed value of $J$,
$L_{\sigma}$ and $S_{6q}$.
 The function $B_{L_{\sigma}LS}^J(k,E)$ is the vertex function in
 the transition $NN \leftrightarrow 6q+\sigma$ and $Z^{JM}_{LS}\in {\cal H}^{\rm ex}$ is
 the transition form factor in the $NN$ channel.

 It is essential that the form of the radial functions $Z^{JM}_{LS}(r)$ can be derived from the
 quark-model calculations in terms of the RGM ansatz. Moreover, the vertex functions
 $B_{L_{\sigma}LS}^J(k,E)$ can be also calculated within the same microscopic
 model~\cite{JPhys2001,IntJModPhys2002}. For the convenience of the reader, in
 Appendices~\ref{formfactor} and \ref{dressing}, we give a brief summary of the main
 assumptions and relationships used in the derivation of the functions
 $Z^{JM}_{LS}(r)$ and $B_{L_{\sigma}LS}^J(k,E)$ in Ref.~\cite{IntJModPhys2002}.

After substituting Eq.~(\ref{tr2}) into Eq.~(\ref{we}), one obtains an
effective potential $w(E)$ induced by coupling the external
$NN$ channel to the internal dibaryon channel in
a form of a sum of simple separable terms for each partial wave:
 \begin{equation}
w(E)=\sum_{S,J,L,L'}V^{SJ}_{LL'}(E),
 \label{nqn2}
 \end{equation}
 with
\begin{equation}
V^{SJ}_{LL^{\prime}}(E)= \sum_M |Z^{JM}_{LS}\rangle\,\lambda^{J}_{SLL^{\prime}}(E)\,
\langle Z^{JM}_{L^{\prime}S}|,
 \label{zlz}
 \end{equation}
where the energy-dependent coupling constants
$\lambda^{J}_{SLL^{\prime}}(E)$ are expressed in terms of the integral over the
momentum ${\bf k}$ of the product of two transition vertices $B$ and the convolution of
the meson and quark propagators:
 \begin{equation}
\lambda^{J}_{SLL^{\prime}}(E)=\int\limits^{\infty}_0\, d{\bf k}
\frac{B_{L_\sigma LS}^J({\bf k})\,{B_{L_\sigma L'S}^J}^*({\bf k})}
{E-E_{\alpha}(k)}.
 \label{lamb}
 \end{equation}
Eqs.~(\ref{nqn2}), (\ref{zlz}) and (\ref{lamb}) are the main result of the
quark microscopic treatment for the $NN$ interaction within the two-component formalism. One can see that
incorporation of the non-nucleon (dibaryon) components leads to an effective $NN$
interaction $w(E)$, which has the form of a sum of separable terms with a specific
energy dependence and which, in particular, can include the coupling between $NN$
channels with different values of the orbital angular momentum $L$, i.e., tensor mixing.

\subsection{A repulsive core effect in the external $NN$ channel}
\label{ortogon}

As has been noted in Sec.~\ref{6qNN}, the important feature of the suggested mechanism for the transition
between the external and internal channels is the presence of two excited $p$-shell quarks
in the incident $NN$ channel, which go to the $s$ shell with emission of two highly coherent
pions (see Appendix \ref{dressing} for details).
Such a mechanism spans the transitions $s^4p^2\to s^6$ and $s^3p^3\to s^5p$ when the inner-channel states are described by the most symmetric configurations
$s^6$ and $s^5p$ in even and odd partial waves, respectively. In line with this assumption, we
should primarily consider in the external $NN$ channel the states which correspond to the quark configurations $s^4p^2$ and $s^3p^3$. One should, first of all, take into account
their orthogonality to the inner states $s^6$ and $s^5p$. According to the results of Sec.~\ref{6qNN} and~\ref{node}, $NN$ wavefunctions in $S$ and some of $P$ partial waves
have a definite nodal structure, which reflects this orthogonality. As was
shown in Refs.~\cite{JPhys2001,IntJModPhys2002}, such a nodal behavior reproduces an effect of the traditional repulsive core at short $NN$ distances\footnote{It should be noted that Neudatchin et
al.~\cite{Neudatchin1975,Neudatchin1977,Obukhovsky1979} were the first to
establish this fact back in the 1970s. In succeeding years,
some quantitative attempts were made to confirm this microscopically in terms of
a constituent quark model~\cite{Kusainov1991} or phenomenologically in terms of
the Moscow $NN$ potential~\cite{Sazonov}.}.

To obtain the correct nodal behavior of the $NN$ wavefunction, it is necessary to
ensure its orthogonality to the corresponding symmetric $6q$ state, using a
projector onto this state, e.g., $P(s^6)$ for the $S$-wave case. In the space of
$NN$ variables, i.e., in the external  channel, it reduces to the one-dimensional
projection operator $P$:
 \begin{equation}
\langle\psi_N\psi_N|P_{\rm sym}|\psi_N\psi_N\rangle\equiv  P
=|\phi_{0}\rangle\langle\phi_{0}|. \label{proj}
\end{equation}
If one uses the harmonic oscillator (h.o.) wave functions for the
nucleon ($s^3[3]$) and the six-quark ($s^6[6]$) states, the form factor
$|\phi_0\rangle$ in the external space is
just the nodeless h.o. $|0s\rangle$ state in the $NN$ relative-motion variable:
 \begin{equation}
|\phi_0\rangle = |0s\rangle_{\rm h.o.}.
\label{phi0}
\end{equation}

Now, to exclude an admixture of the $0s$
function $|\phi_0\rangle$ and to ensure the presence of a stationary node in the external-channel wavefunctions, as required by the microscopic treatment (see, e.g., Eqs.~(\ref{q7}) and (\ref{q7prime})), the Schr\"odinger equation (\ref{Sch2}) for the wavefunction in the external channel should be solved with an additional orthogonality constraint:
\begin{equation}
\langle\Psi^{\rm ex}|\phi_0\rangle =0.
\label{orth}
\end{equation}
To solve equations with an additional orthogonality condition like (\ref{orth}), it is convenient to use the orthogonal projection method (see, e.g.,~\cite{Sazonov}). In the method, the equation is solved in full space, but a projector onto the subspace to be excluded [the projector $P$ (\ref{proj}) in our case] with a large coupling constant $\lambda_0$ is added to the original Hamiltonian.
Thus, we obtain the final form of the effective equation for the external-channel wavefunction:
\begin{equation}
(h^{\rm ex}+ w(E) +\lambda_0P -E)\Psi^{\rm ex} = 0. \qquad
\label{effeq}
\end{equation}
The orthogonalizing term $\lambda_0P$ in the effective equation (\ref{effeq}) replaces the traditional repulsive core in the $NN$ interaction.
It should be noted that although the orthogonalizing term $\lambda_0 P$ can be formally assigned to the external channel and included in the external Hamiltonian $h^{\rm ex}$, its appearance is associated with the six-quark symmetry of the system, i.e., with non-nucleonic degrees of freedom.

A similar situation takes place also in $P$ waves. Here the $6q$ wavefunction of the cluster $NN$ channel should not contain an admixture of the $s^5p[51]_X$ configuration, which is associated with a quark core.
Therefore, passing to the variables of $NN$ relative motion, one gets the orthogonality condition:
\begin{equation}
\langle\Psi^{\rm ex}|\phi_1\rangle =0,\, |\phi_1\rangle = |1p\rangle_{\rm h.o.}
\label{orth1}
\end{equation}
and the corresponding projector.

Formally, it is necessary to take a limit $\lambda_0\to \infty$ to completely exclude the most symmetric $6q$ configurations from the external channel. In the practical calculations, it is usually sufficient to take the value of $\lambda_0$ to be ca. $10^5$--$10^6$~MeV. Note that the admixture of excluded states decreases with increasing $\lambda_0$ as $\lambda_0^{-1}$.
However, keeping in mind that there are no completely forbidden states in the $6q$ system, we can consider a more general case when the admixture of the symmetric component in the wavefunction is not completely excluded, but limited. This can be done using
the same projector $P$ in the external channel with a finite value of $\lam_0$ ca. $10^2$--$10^3$~MeV. Such a form of interaction is employed below in Sec.~\ref{sec_recent}.

\subsection{The results of calculations with the dibaryon (dressed bag) model}

The version of the dibaryon model described above, based on the microscopic six-quark description of the $NN$ system and referred to as the dressed bag model (DBM)~\cite{JPhys2001,IntJModPhys2002}, results in the equation (\ref{effeq}) with the following effective $NN$ interaction:
\begin{equation}
V^{\rm eff}=v^{\rm ex}+ w(E) +\lambda_0P.
\label{Veff}
\end{equation}
In the DBM, the interaction in the external channel $v^{\rm ex}$ included the one-pion exchange potential (OPEP) with a soft dipole cutoff:
\begin{equation}
v^{\rm OPE}=-
\frac{f_{\pi}^2}{m_{\pi}^2}\frac{({\bm\tau}_1 {\bm\tau}_2)}{3}\frac{({\bm\sigma}_1 {\bf q})({\bm
\sigma}_2{\bf q})}{q^2+m_{\pi}^2}\left(\frac{\Lambda_{\pi
NN}^2-m_{\pi}^2}{\Lambda_{\pi
NN}^2+q^2}\right)^2,
\label{Vope}
\end{equation}
with $m_{\pi} = (m_{\pi^0} + 2m_{\pi^{\pm}})/3$ being the averaged
pion mass, $f_{\pi}^2/(4\pi)=0.075$ the averaged pion-nucleon coupling
constant and $\Lambda_{\pi NN} \simeq 0.6$--$0.7$~GeV/$c$ the high-momentum cutoff
parameter,
and a small potential $v^{\rm TPE}$ which provided an additional attraction (ca. $2$-–$3$~MeV)
in the region $r \sim 1.5$--$2$~fm:
 \begin{equation}
 v^{\rm TPE}(r)=v_0^{\rm TPE}\,(\beta r^2)^2\exp(-\beta r^2).
\label{TPE}
\end{equation}
In Ref.~\cite{IntJModPhys2002} it was assumed that such a potential could represent the contribution of the peripheral part of the two-pion exchange. This contribution turned out to be important for a precise description of the scattering length and effective radius in the $^1S_0$ and $^3SD_1$ channels. It should be noted however that in the modified version of the dibaryon model generalized to higher partial waves (see Sec.~\ref{sec_recent}), the external interaction $v^{\rm ex}$ does not include the terms similar to the potential $v^{\rm TPE}$.

The model was employed in Ref.~\cite{IntJModPhys2002} for the description of $NN$ elastic scattering in the $^1S_0$ and $^3SD_1$ partial waves and the deuteron properties.
A very good description for the elastic scattering data in the energy region from zero to 1 GeV including the low-energy parameters (scattering length and effective range) and the deuteron static properties such as quadrupole momentum, charge radius and others, was obtained with the weight of the internal (dibaryon) component in the deuteron ca. 3.6\%.

In a system of several nucleons, each pair of nucleons can form an intermediate dibaryon state. Consequently, the dibaryon concept inevitably leads to the emergence of a new three-body force, which arises due to the interaction of a dressed dibaryon formed from a given pair of nucleons with another (third) nucleon.
 Applying the dibaryon model for $NN$ and $3N$ interactions in a three-nucleon system resulted in a good description of the binding energies for the $^3$H and $^3$He nuclei and their Coulomb difference~\cite{sys3n,YAF3N}. The DBM was also tested in the calculations of the ground states of the $^6$Li and $^6$He nuclei~\cite{6Li} within the framework of the three-cluster $\alpha+2N$ model, with account of a new three-particle force induced by the interaction of the internal dibaryon state with the $\alpha$ core.

Recently, we proposed~\cite{YAF19,PLB20,EPJA20,PRD20} a modified version of the dibaryon model generalized to higher partial waves, which takes into account the presence of experimentally detected dibaryon resonances and allows one to describe both elastic and inelastic $NN$ scattering in a broad energy range well above the pion production threshold.
This version of the model is considered in the next Sections.

\section{Description of elastic and inelastic $\bm{NN}$ scattering}
\label{sec_recent}

One can make some further simplification of the dibaryon model when the internal
space consists of a single state $|\alpha\rangle$ only and the meson degrees of
freedom in the internal channel are not considered explicitly
\cite{PLB20,PRD20}. At the same time, the probability of coupling between this
state and its possible non-nucleonic decay channels can be taken into account
explicitly. In this approximate treatment, the effective interaction $w(E)$ in
the external channel has a pole-like energy dependence, however, the pole
position has an imaginary part that corresponds to the possible decays of the
internal $6q$ state into inelastic (non-nucleonic) channels. This form of
interaction allows one to consider both elastic and inelastic processes in $NN$
scattering. Also, this effective interaction leads to a presence of resonances
in the whole system, the positions of which can be compared with experimental
data. One can expect that such a single-pole approximation for the effective
interaction $w(E)$ is justified primarily for the higher $NN$ partial waves
where the coupling constants between external and internal channels should be
rather small. However, as it has been shown in Ref.~\cite{EPJA20}, the case of
strong coupling which takes place in $S$ waves can be also considered within
this approximation.

Below we briefly summarize the main results obtained for this version of the model.

\subsection{The effective interaction}
\label{form}

With the above simplifications, the external Hamiltonian has the same form as in Eq.~(\ref{effeq}). It consists of three terms, i.e., the kinetic energy, the OPEP\footnote{In our present calculations, the soft cutoff
$\Lambda_{\pi NN}=0.65$~GeV/$c$ is used for all the $NN$ channels except for the $^1S_0$ and coupled $^3SD_1$ configurations where
a bit smaller value 0.62~GeV/$c$ is taken \cite{EPJA20}.} (\ref{Vope}) and the repulsive
orthogonalizing potential for some partial waves.
Here the value of $\lam_0$ in the last term keeps to be finite which corresponds
to an incomplete (partial) exclusion of the symmetric $6q$ configuration from the external channel.

The energy-dependent interaction takes a pole-like form:
\begin{equation}
w(E)= \frac{|Z\rangle \langle
Z|}{E-E_D}, \label{Vdib}
\end{equation}
where $E_D$ is the pole position (see below) and $|Z\rangle$ --- the transition form factor which includes the coupling strengths $\mu$. For the singlet $NN$ partial waves with a definite value of the orbital momentum $L$, one has $|Z\rangle=\mu_L|\phi_L\rangle$.
For the coupled spin-triplet channels with the total angular momentum $J$
and the tensor coupling of states with orbital
momenta $L=J-1$ and $J+1$, we still consider a single state in
the internal subspace which, however, couples to both partial
external channels. In this case, the transition form factor has the following two-component form:
$|Z\rangle\equiv \left(\begin{array}{c}
\mu_{J-1}|\phi_{J-1}\rangle\\
\mu_{J+1}|\phi_{J+1}\rangle\\
\end{array}\right)$. Thus, for both
singlet and triplet $NN$ partial channels, the structure of the
interaction potential is the same. The coupling constants from Eq.~(\ref{lamb}) have a simple pole-like energy dependence:
\begin{equation}
\lam^{J}_{SLL'}(E)=\frac{\mu_L\mu_{L'}}{E-E_D},
\end{equation}
where the values of the partial strengths $\mu_L$, $\mu_{L'}$ and the energy $E_D$ depend on the total angular momentum $J$ and spin $S$.

Similarly to the DBM treatment, the form factors $|\phi_0\rangle$ and $|\phi_L\rangle$ are taken as the
h.o. functions with the same scale parameter
$r_0$, in accordance to the shell model:
\begin{eqnarray}
\phi_{0L}(k)=A_{0L}(kr_0)^{L+1}e^{-\half(kr_0)^2},\label{nlr}\\
\phi_L(k)=A_{1L}(kr_0)^{L+1}
\left[L+\tralf-(kr_0)^2\right]e^{-\half(kr_0)^2},\label{nlr1}
\end{eqnarray}
where $A_{0L}$ and $A_{1L}$ are the normalization factors.
If the potential contains the orthogonalizing term with the
nodeless function (\ref{nlr}), then the form factor in the coupling
term has the form (\ref{nlr1}) with {\em the same} parameter $r_0$. If the orthogonalizing term
is absent (i.e. $\lam_0=0$), the coupling form factor is taken in the nodeless form
(\ref{nlr}).

Thus, in this version of the model, the general structure of the interaction remains the same as in the DBM. The main difference is related to the energy dependence of the coupling constants (\ref{lamb}).

\subsection{Account of inelastic processes}
\label{inel}

For the effective account of inelastic processes and the description
of the threshold behavior of the reaction cross section in different partial waves, we
may consider the complex-valued energy $E_D=E_0 -i\Gamma_{D}/2$, which corresponds to a ``bare'' dibaryon resonance\footnote{Actually, we mean here a $6q$ state dressed with meson loops, but not with the $NN$ loops.}, and introduce further the energy dependence of the width $\Gamma_D$. Here we assume that inelastic processes occur via the corresponding dibaryon resonance decay. For example, one-pion production goes via the decay modes $D \to \pi N N$ and $D \to \pi d$. The first mode is actually the dominant one for the known isovector dibaryons~\cite{Strak91}. In general, the decay widths for both these modes should have similar threshold behaviour, so that, for simplicity, we include just the first one in the $\Gamma_{D}$ parametrization and adjust the parameters to effectively take into account the total inelastic width. Thus, we adopt the following representation of $\Gamma_{D}$:
\begin{equation}
\Gamma_D(\sqrt{s})=\left\{
\begin{array}{lr}
0,& \sqrt{s}\leq E_{\rm thr};\\\displaystyle
\Gamma_0\frac{F(\sqrt{s})}{F(M_0)},&\sqrt{s}>E_{\rm thr}\\
\end{array}\label{gamd}
\right\},
\end{equation}
where $\sqrt{s}$ is the total invariant energy of the decaying
resonance, $M_0$ --- the ``bare'' dibaryon mass, $E_{\rm
thr}=2m+m_\pi$ --- the threshold energy, and $\Gamma_0$ defines
the decay width at the resonance energy.

The function $F(\sqrt{s})$ should take into account the dibaryon
decay into the $\pi N N$ channel. So that, for the given values of
the orbital angular momenta of the pion $l_{\pi}$ and $NN$ pair
$L_{NN}$, this function can be parameterized as follows:
\begin{equation}
F(\sqrt{s})=\frac{1}{s}\int_{2m}^{\sqrt{s}-m_{\pi}}dM_{NN}
\frac{q^{2l_\pi+1}k^{2L_{NN}+1}}{(q^2+\Lam^2)^{l_\pi+1}(k^2+\Lam^2)^{L_{NN}+1}},
\label{fpinn}
\end{equation}
where $\displaystyle
q={\sqrt{(s-m^2_\pi-M^2_{NN})^2-4m_\pi^2M_{NN}^2}}\Big/{2\sqrt{s}}$
is the pion momentum in the total c.m.s.,
$\displaystyle k=\half\sqrt{M_{NN}^2-4m^2}$ --- the momentum of
the nucleon in the c.m.s. of the final $NN$
subsystem with the invariant mass $M_{NN}$, and $\Lam$ --- the
high-momentum cutoff parameter which prevents an unphysical rise
of the width $\Gamma_{D}$ at high energies. The orbital
momenta $l_{\pi}$ and $L_{NN}$ may take different values however
their sum is restricted by the total angular momentum and parity
conservation.

In the isoscalar channels, the main inelastic process is
two-pion production. In this case, one may use Eq.~(\ref{gamd}) and an expression for $F(\sqrt{s})$ similar to Eq.~(\ref{fpinn}) but for the $D\to \pi\pi d$ decay~\cite{PLB20}.

The separable form of the interaction~(\ref{Vdib}) allows one to find
the resonance parameters straightforwardly. In this case, the explicit expressions for the $S$-matrix and inelastic cross section can be written. The latter has a form which is similar to the Breit--Wigner (BW) one. However, it contains an additional energy dependence in both terms of the  BW denominator. So that, the resulting energy-dependent width consists of the initial ''bare'' width $\Gamma_D$ and a term which results from the coupling with the external $NN$ channel. The resonance position occurs to be shifted due to this coupling as well (see details in Refs.~\cite{YAF19} and~\cite{PLB20}).
Thus, in this version of the model, the coupling between the
external $NN$ (driven by OPEP) channel and the internal (``bare''
dibaryon) channel leads to a renormalization of the complex
energy of the initial ``bare'' dibaryon and its transformation to
the physical mass and width of the ``dressed'' dibaryon which can be deduced from experimental data.


\subsection{Elastic and inelastic $NN$ scattering amplitudes}
\label{ampl}

The results for particular $NN$ partial-wave amplitudes have been reported in
Refs.~\cite{YAF19,PLB20,EPJA20,PRD20}. Here we summarize the results for the
lowest 14 $NN$ partial channels, including new and updated fits for some channels which have not been considered previously.
We use the conventional $K$-matrix notations~\cite{Arndt} for the partial phase shifts,
inelasticity parameters and mixing angles.

\subsubsection{Channels related to the known dibaryon resonances}
\begin{figure*}[t]
\centering\epsfig{file=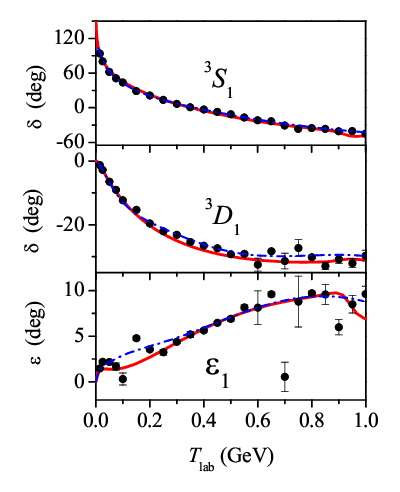,width=0.6\columnwidth}
       \epsfig{file=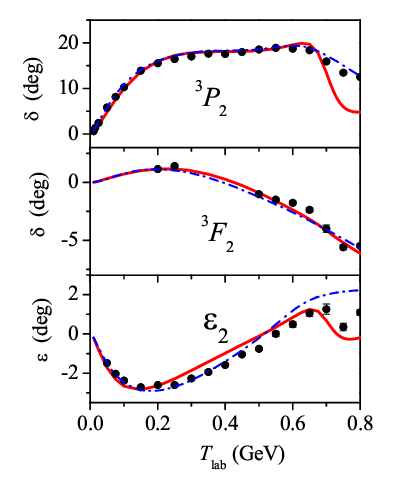,width=0.6\columnwidth}
       \epsfig{file=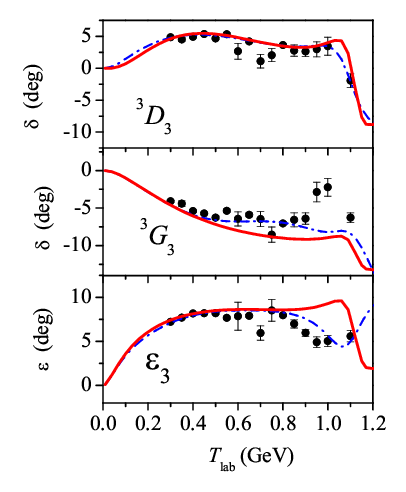,width=0.6\columnwidth}
 \caption{\label{fig_coupled} Phase shifts and mixing angles for the coupled $NN$ partial channels found within the dibaryon model (solid curves)
   in comparison with the SAID PWA single-energy (circles) and energy-dependent (dash-dotted curves) solutions: SM16~\cite{SAID} for the $^3S_1$--$^3D_1$ and $^3P_2$--$^3F_2$ channels and AD14~\cite{AD14} for the $^3D_3$--$^3G_3$ channels.}
\end{figure*}

 We start from 10 $NN$ channels (including three coupled-channel configurations) where experimental evidences for the
 corresponding 7 dibaryon resonances take place (see Sec.~\ref{sec_status}). The model parameters for these channels are collected in Tab.~\ref{Tab1}. The data for the channels with definite values of $J$ are separated by lines.
\begin{table}[h!]
\caption{The dibaryon model parameters for the lowest
$NN$ partial channels.\label{Tab1}}
\begin{center}\begin{tabular}{c c c c c }\hline  $^{2S+1}L_J$ &
$\lam_0$ (MeV)&$r_{0}$ (fm)&$\mu_{L}$ (MeV) & $M_0$ (MeV) \\
\hline
$^1S_0$ & 165 &0.48 & 274.2& 2300.313  \\
\hline
$^3S_1$& 165 & 0.475 & 248.1& 2275.69  \\
$^3D_1$& 0 & 0.6 & 65.9&  \\
\hline
$^3P_0$ & 450 & 0.425 & 35 &  2200 \\
\hline
$^3P_2$ & 0 & 0.7 & 65 &  2205 \\
$^3F_2$ & 105 & 0.45 & 1.5 &  \\
\hline
$^1D_2$ & 0 & 0.82 & 48 &  2168 \\
\hline
$^3D_3$ & 0 & 0.71 & 58 &  2363 \\
$^3G_3$ & 0 & 0.71 & 36 &   \\
\hline
$^3F_3$ & 0 & 0.5 & 70 &  2240 \\
 \hline
\end{tabular}\end{center}
\end{table}
The parameter values have been fitted
to reproduce the partial phase shifts and inelasticities in each partial configuration.
 Thus, we obtained the ``bare'' masses $M_0$ to be rather close to the masses of the known dibaryon resonances which, in turn, are close to the respective di-hadron thresholds ($N\Delta$, $NN^*(1440)$ or $\Delta\Delta$) as discussed in Sec.~\ref{sec-thr}. The values of $\lam_0$ reflect the repulsive part of the interaction which is also consistent with the nodal structure of the corresponding wavefunctions (see Sec.~\ref{node}). The values of $r_0$ should depend on the features of the internal state as well. In fact, $r_0$ changes not so arbitrary as it may seem from Tab.~\ref{Tab1}. As it has been noted above, we use different form factors given by Eqs.~(\ref{nlr}) and (\ref{nlr1}), i.e., represented by the nodeless h.o. function (when $\lam_0=0$) and the h.o. function with one node (when $\lam_0\neq0$), respectively. The effective ranges $R=\sqrt{\langle r^2\rangle}$ of the above h.o. functions are different and additionally depend on the orbital momentum value $L$. Thus, $R=r_0 \sqrt{L+3/2}$ for the nodeless function and $R=r_0 \sqrt{L+7/2}$ for the h.o. function with one node. Taking into account these effective ranges, one can find that the smallest $R \approx 0.9$ fm corresponds to the $S$ waves and $^3P_0$ wave, and the value of $R$ rises for higher partial waves, which looks reasonable\footnote{A more detailed analyses of the model parameters will be given in the forthcoming study.}.

\begin{figure}[h]
\centering\epsfig{file=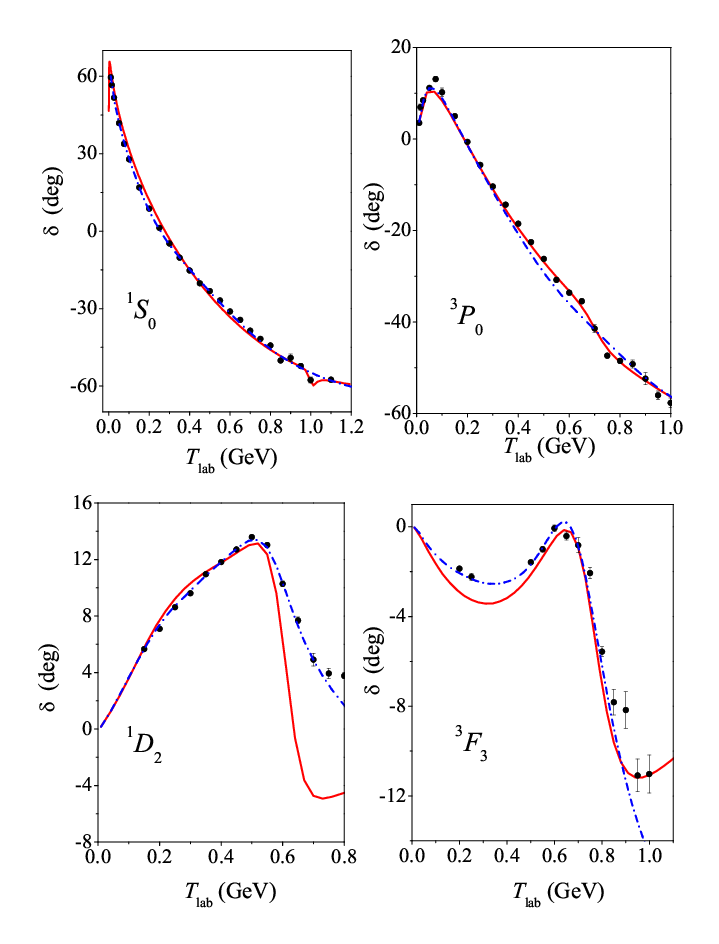,width=0.9\columnwidth}
 \caption{\label{4waves} Phase shifts for the uncoupled $NN$ partial channels $^1S_0$, $^3P_0$, $^1D_2$ and $^3F_3$ found within the dibaryon model (solid curves) in comparison with the SAID
 PWA single-energy (circles) and energy-dependent SM16 (dash-dotted curves) solutions~\cite{SAID}.}
\end{figure}

In Fig.~\ref{fig_coupled}, the partial phase shifts, mixing angles
and inelasticity parameters for the coupled spin-triplet channels
$^3S_1$--$^3D_1$, $^3P_2$--$^3F_2$ and $^3D_3$--$^3G_3$ are shown in comparison with the single-energy and energy-dependent solutions of the SAID PWA \cite{SAID,AD14}. We compare all our results with the SAID solution SM16 \cite{SAID}, except for the $^3D_3$--$^3G_3$ channels, where we use for comparison the solution AD14 \cite{AD14}, which gives an $S$-matrix pole corresponding to the $d^*(2380)$ dibaryon resonance.
 The partial phase shifts for the rest uncoupled channels $^1S_0$, $^3P_0$, $^1D_2$ and $^3F_3$ are presented in Fig.~\ref{4waves}.

One can see quite good agreement between the dibaryon model predictions and the SAID PWA solutions up to rather high laboratory energies $T_{\rm lab}$, which correspond to the resonance position in each case.
For the coupled-channel configurations, some discrepancies are seen for the $^3G_3$ phase shift and mixing angle $\epsilon_3$ which require a further improvement of the model. At the same time, the coupled $NN$ channels $^3D_3$--$^3G_3$ correspond to the most evident case, for which the resonance-like behavior of the $NN$ elastic scattering amplitudes (due to the dibaryon state $d^*$(2380)) has been found experimentally and confirmed by the SAID PWA as well~\cite{AD14}.

Our model description of the $NN$ scattering amplitudes also takes into account inelastic processes.
\begin{figure*}[t]
\centering\epsfig{file=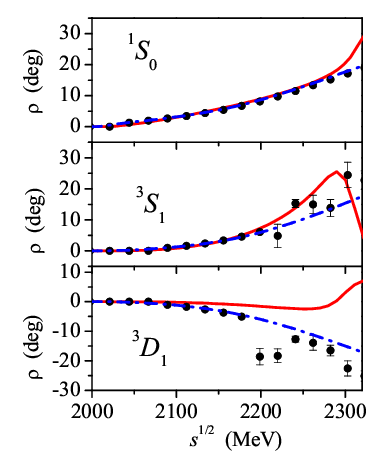,width=0.6\columnwidth}
       \epsfig{file=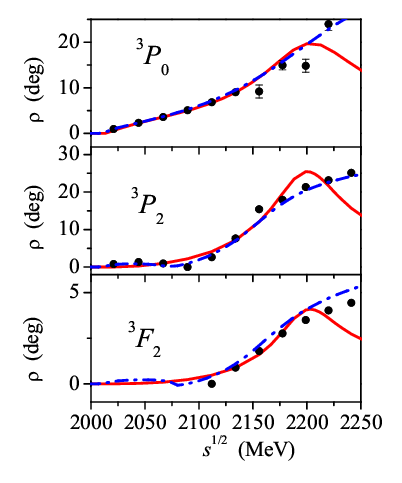,width=0.6\columnwidth}
       \epsfig{file=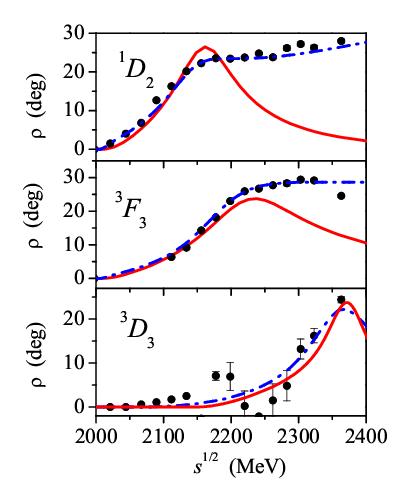,width=0.6\columnwidth}
 \caption{\label{inel123} Inelasticity parameters for the coupled and uncoupled $NN$ partial channels 
 found within the dibaryon model (solid curves) in comparison with the SAID
 PWA single-energy (circles) and energy-dependent (dash-dotted curves) solutions~\cite{SAID,AD14}.}
\end{figure*}
For the majority of partial channels, the parametrization (\ref{gamd})
for the imaginary part of the internal-state
energy $E_D$ with a decay width in the form of Eq.~(\ref{fpinn}) corresponding to the $D \to \pi N N$ decay has been used.
 The respective parameter values are listed in Tab.~\ref{Tab2}. Here, the values of $\Gamma_0$ differ evidently which is caused particularly by a strong dependence of the pion production probability on the $NN$ channel quantum numbers. The available parameters $l_{\pi}$ and
$L_{NN}$ are restricted by the parity and total angular momentum conservation, and their particular values, as well as the cut-off parameter $\Lam$, have been adjusted to get the best fit of the SAID single-energy data for the inelasticity $\rho$ in the near-threshold region. The large spread of the $\Lam$ values indicates the need of a more accurate treatment of the dibaryon width in some partial channels. In particular, a very slow rise of the $^3S_1$ inelasticity from threshold is related to the low single-pion production probability and importance of the two-pion decay mode in this channel. Since we focus here on the general description of $NN$ data in a wide energy range, we postpone the refinement of the model for a better description of the near-threshold behavior of some partial amplitudes to a future work.

\begin{table}[h!]
\caption{Parameters used for the representation of $\Gamma_D$ (see Eqs.~(\ref{gamd}) and~(\ref{fpinn})) in the lowest $NN$ partial channels.\label{Tab2}}
\begin{center}\begin{tabular}{c c c c c }\hline  $^{2S+1}L_J$ &
$\Gamma_0$ (MeV)& $\Lam$ (MeV)& $l_{NN}$ & $l_{\pi}$  \\
\hline
$^1S_0$ & 40 & 300 & 0 & 1  \\
$^3SD_1$&  80 & 1800 & 2 & 1 \\
$^3P_0$ & 92 & 87 & 0 &  0 \\
$^3PF_2$ & 100 & 1000 & 0 &  2 \\
$^1D_2$ & 100 & 300 & 1 &  0 \\
$^3F_3$ & 150 & 100 & 2 &  0 \\
 \hline
\end{tabular}\end{center}
\end{table}

For the coupled channels $^3DG_3$, the dibaryon width (\ref{gamd}) should include predominantly the two-pion decay mode, which gives the main impact to the inelasticity here. So, we used here the parametrization of the decay width similar to Eq.~(\ref{fpinn}) but for the $D \to
d\pi\pi$ decay, with the following parameters: $\Gamma_0=60$ MeV, $l_d=l_{\pi\pi}=1$, $\Lam=150$ MeV
(see details in Ref.~\cite{PLB20}).

The comparison of inelasticity parameters $\rho$ with the SAID PWA data for the majority of the $NN$ partial channels is shown in Fig.~\ref{inel123}.
For the visual representation, the dependence of the data on the total invariant energy $\sqrt{s}$ is used here.
Except for the $^3D_1$ partial wave, all the figures reflect good agreement
of the model calculations with the PWA data up to the invariant energy values corresponding to the pole
position. At higher energies, the calculated inelasticities decrease,
while the PWA data still grow up. This behavior has been expected,
because we employ the resonance-like parametrization for the decay
width. Other inelastic processes not included into our model treatment should also give impact to the total inelastic amplitudes at higher energies.

Finally, in Tab.~\ref{Tab3}, the dressed
dibaryon resonance parameters found from our model fits are compared with the parameters obtained by PWA or phenomenological analysis from the experimental data\footnote{The ``trivial'' dibaryons in the $^3SD_1$ (the deuteron) and
$^1S_0$ (the singlet deuteron) partial channels are not included here.}. The raws of the Table are sorted by the increasing resonance masses.
Nearly all the dibaryon parameters occur to be in quite reasonable agreement with the empirical values quoted in the last column of the Table.
\begin{table}[h!]
\caption{Parameters ($M_{\rm th}$, $\Gamma_{\rm th}$) of the
dressed dibaryons (in GeV) found in the present model for seven $NN$
configurations and their empirical values from the references given in
the last column.\label{Tab3}}
\begin{center}\begin{tabular}{c c c c c c c }\hline
$\displaystyle {}^{2S+1}L_J$
& $T(J^P)$ & $M_{\rm th}$&$\Gamma_{\rm th}$&$M_{\rm exp}$&$\Gamma_{\rm exp}$ & Ref.\\
\hline
$^1D_2$& $1(2^+)$ & 2.18&0.14& 2.14--2.18 &0.05--0.11 & \cite{Hoshiz,Strak91}\\
$^3P_0$& $1(0^-)$ & 2.2 & 0.099 & 2.201(5) & 0.091(12) & \cite{Komarov16}\\
$^3PF_2$& $1(2^-)$ & 2.221& 0.17 & 2.197(8) & 0.130(21) & \cite{Komarov16}\\
$^3F_3$& $1(3^-)$ & 2.23&0.185&  2.20--2.26 & 0.1--0.2 & \cite{Hoshiz,Strak91}\\
$^3SD_1$&  $0(1^+)$ & 2.31& 0.16 & 2.315(10) & 0.150(30) & \cite{EPJA20}\\
$^1S_0$& $1(0^+)$ & 2.33& 0.05 & 2.32 & 0.15 & \cite{EPJA20}\\
$^3DG_3$ & $0(3^+)$ & 2.376 & 0.084 & 2.38(1) & 0.08(1) & \cite{AD14}\\
\hline
\end{tabular}\end{center}
\end{table}
The resulting shift of the dressed dibaryon mass $M_{\rm th}$ with regard to the initial position $M_0$ given in Tab.~\ref{Tab1} for each partial configuration is caused by the coupling between the internal and external ($NN$) channels and depends on the coupling strength $\mu_L$, the type of the form factor and the external part of the $NN$ interaction (see details in Refs.~\cite{YAF19,PLB20}).

The results presented in Figs.~\ref{fig_coupled}--\ref{inel123} show that the pole-like form of the interaction
leads to a rather good description of the $NN$ partial channels with an evident repulsion. In particular, the phase shifts are reproduced quite well even at energies above the resonance positions, though the inelasticities have the wrong decreasing there. At the same time, for the channels without repulsion, such as $^1D_2$, $^3P_2$ and $^3DG_3$, the phase shifts are described properly up to the resonance energies only. This behavior is directly related to the decrease of inelasticities. Thus, some refinement of the model is required if one needs to move to the higher energy region. This could be done by an inclusion of other possible inelastic processes, account for the dynamics in the internal channel, etc.

To summarize, we have achieved a very good quantitative
description for the partial phase shifts in the $NN$ channels
considered and a rather good qualitative description for the
inelasticity parameters in a broad energy
range starting from zero energy which is very far from the
position of the ``bare'' dibaryon. The very important feature of the
results presented above is good agreement of the masses and widths
of the dressed dibaryons obtained by fitting the $NN$ phase shifts and inelasticities
in our model with the dibaryon parameters deduced from experiments.

The success of the model for the $NN$ channels with the known dibaryon resonances
allows us to make a step forward and extend the suggested model for the partial channels where
the existence of dibaryon resonances is not confirmed to date.

\subsubsection{Partial $NN$ channels for which dibaryon states have not been found yet}
Below we discuss the uncoupled $NN$ partial channels where
the dibaryon resonances have not been reliably detected to date. These are the isovector
channel $^3P_1$ and three isoscalar channels $^1P_1$, $^3D_2$ and $^1F_3$.

For some of the $NN$ channels in question, indications of possible resonances exist in the literature. In particular, the $^3P_1$ resonance was found in the PWA~\cite{Strak}. We have constructed the dibaryon potential for the $^3P_1$ channel in Ref.~\cite{PRD20}, where the resonance with a position rather close to those for the $^3P_0$ and $^3P_2$ channels has been predicted. For the isoscalar $NN$ channel $^1P_1$, the estimations can be done based on the resonance-like behavior of the
isoscalar part of the $NN$-induced single-pion production cross section near the $NN^*(1440)$ threshold~\cite{Clem20}. At the same time, we are not aware of any indications of the dibaryon resonances in the channels $^3D_2$ and $^1F_3$.

The dibaryon model parameters used for fitting the partial phase shifts in the above $NN$ channels are collected in Tab.~\ref{Tab4}.
 \begin{table}[h!]
\caption{Dibaryon model parameters for the $NN$ partial channels $^1P_1$, $^3P_1$, $^3D_2$ and
$^1F_3$.\label{Tab4}}
\begin{center}\begin{tabular}{c c c c c }\hline  $^{2S+1}L_J$ &
$\lam_0$ (MeV)&$r_0$ (fm)&$\mu_{L}$ (MeV) & $M_0$ (MeV) \\
\hline
$^3P_1$ & 270 &0.425 & 20 & 2230   \\
$^1P_1$ & 280 &0.48 & 90 & 2320  \\
$^3D_2$ & 120 &0.5 & 165 & 2350    \\
$^1F_3$ & 120 &0.51 & 70 & 2345   \\
 \hline
\end{tabular}\end{center}
\end{table}
Here the values of the parameters $M_0$ and $r_0$ for the channel $^3P_1$ occurred to be rather close to those for the channel $^3P_0$ (see Tab.~\ref{Tab1}), while for other three (isoscalar) channels, they are comparable with the values for the $^3S_1$ channel.

We have also fitted the inelasticity parameters for the partial
channels $^3P_1$ and $^1P_1$. The dibaryon width has been parameterized
by Eqs.~(\ref{gamd}) and (\ref{fpinn}). The respective parameters are given in
Tab.~\ref{Tab5}.
\begin{table}[h!]
\caption{Parameters of the dibaryon width used in the partial
channels $^3P_1$ and $^1P_1$.\label{Tab5}}
\begin{center}\begin{tabular}{c c c c c}\hline  $^{2S+1}L_J$ &
$\Gamma_0$ (MeV)& $\Lam$ (MeV)& $l_{NN}$ & $l_{\pi}$  \\
\hline
$^3P_1$ & 50 & 200 & 0 &  0 \\
$^1P_1$ & 110 & 300 & 1 &  1 \\
 \hline
\end{tabular}\end{center}
\end{table}
We have not considered the inelasticities in the $^3D_2$ and $^1F_3$ partial waves, since they
are missing in the SAID PWA. For these partial channels, the invariant mass
of the resonance can be definitely found, while the width can
not be estimated from the fit of elastic phase shifts only. At the same time, the model potential for these channels
has a repulsive part. So that, the decay widths of the internal states contribute to the phase shifts in the narrow region near the resonance position only.

\begin{figure}[h]
\centering\epsfig{file=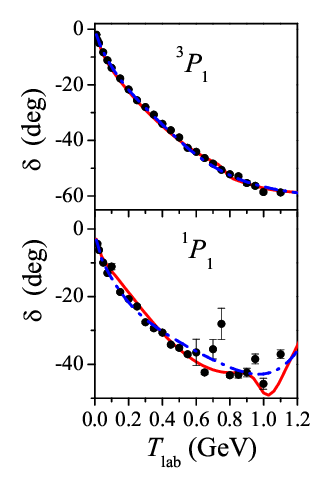,width=0.49\columnwidth}
\epsfig{file=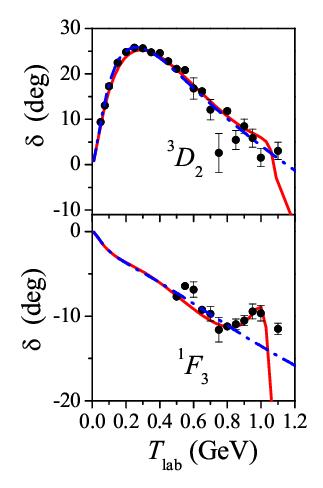,width=0.49\columnwidth}
 \caption{\label{4waves_next} Partial phase shifts for the $NN$ channels $^1P_1$, $^3P_1$, $^3D_2$ and
$^1F_3$ found within the dibaryon model (solid curves) in comparison with the SAID PWA single-energy (circles) and energy-dependent SM16 (dash-dotted curves) solutions~\cite{SAID}.}
\end{figure}

The comparison of the calculated partial phase shifts with the SAID PWA solutions for the above $NN$ channels is shown in
Fig.~\ref{4waves_next}.

\begin{figure}[h]
\centering\epsfig{file=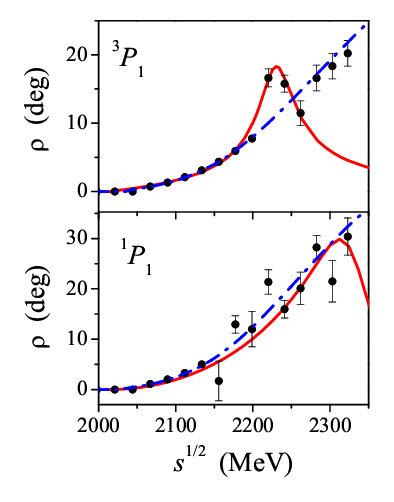,width=0.7\columnwidth}
 \caption{\label{inel_2waves} Inelasticity parameters for the $NN$ partial channels $^1P_1$ and $^3P_1$ found within the dibaryon model (solid curves) in comparison with the SAID PWA single-energy (circles) and energy-dependent SM16 (dash-dotted curves) solutions~\cite{SAID}.}
\end{figure}

The inelasticity parameters for the $P$-wave channels
are presented in Fig.~\ref{inel_2waves}. One can see quite good agreement of our model calculations with the PWA data.

Our theoretical calculations within the dibaryon model predict two
resonances in the $NN$ partial channels $^3P_1$ and $^1P_1$ with the quantum numbers $T(J^P) = 1(1^-)$ and $0(1^-)$, respectively, and the following parameters:
$M_{\rm th}({^3P_1})=2.23$ GeV, $\Gamma_{\rm th} ({^3P_1})=0.05$ GeV and
$M_{\rm th}({^1P_1})=2.33$ GeV, $\Gamma_{\rm th} ({^1P_1})=0.13$ GeV.
We have also predicted two more isoscalar states with the masses close to 2.35 GeV in the
$NN$ channels $^3D_2$ and $^1F_3$, however, their widths could not be defined accurately.

From the results presented in Tab.~\ref{Tab3} and this subsection, one may conclude that dibaryon states can be divided into two groups according to their masses. The majority of the isovector states (except for the $^1S_0$ one) lie near the $N\Delta$ threshold, while the masses of the  isoscalar states (and the $^1S_0$ state as well) are close to the $NN^*$(1440) and $\Delta\Delta$ thresholds. These findings should correspond to the internal structure of the six-quark systems with different isospins. It seems likely that such a dibaryon $|DB\rangle$ can be represented as a mixed state of the six-quark core and the hadron molecule (a loosely bound $N+N^*$ state), $|DB\rangle=\cos\theta|6q\rangle+\sin\theta|NN^*\rangle$, where $\theta$ is the mixing angle. Recently such an approach has been successfully used in the baryon sector for the description of the Roper resonance represented as $|R\rangle=\cos\theta_r|3q\rangle+\sin\theta_r|N\sigma\rangle$~\cite{Obukh19}. The molecular $N\sigma$ component plays a role for the understanding of both the helicity amplitudes of the Roper resonance electroproduction and a large $2\pi$ branching in this reaction.
In the case of the $DB$, there is a good probability that the hadronic molecular mode $NN^*$, when present, will be also seen in the pion electroproduction off the deuteron at large momentum transfer or in the pion photoproduction near the $NN^*$ threshold\footnote{The $NN^*$ mode should contain also the $NN\sigma$ (or dibaryon+$\sigma$) component in the case when $N^*$ is the Roper resonance.}.


\section{Conclusion}
\label{sec_concl}

In this work we presented the new QCD-motivated approach to $NN$ scattering at intermediate energies --- the dibaryon model, which includes the dibaryon resonance formation at short $NN$ distances supplemented by the peripheral one-pion-exchange at long distances. This novel approach implements the duality principle for $NN$ scattering (see., e.g.,~\cite{dual}), i.e., it replaces the  $t$-channel multi-meson exchanges in the traditional $NN$-potential models by the $s$-channel mechanism corresponding to the exchange of the dibaryon resonance (the $6q$ bag dressed with meson fields) between the interacting nucleons in their overlap region. We argued that dibaryon resonances can serve as a bridge between the traditional picture of the $NN$ interaction dealing with point-like nucleons and mesons and the QCD world dealing with quarks and gluons. Based on the microscopic six-quark consideration of the $NN$ system, we have shown that the dibaryon degrees of freedom are appropriate to effectively take into account the inner structure of the nucleons in the $NN$ scattering processes.

    The purpose of this work was not to prove the existence of dibaryon resonances. Instead, we employed their parameters deduced from analysis of the recent experiments~\cite{Clem17,Clem21} to build some alternative picture of the short-range $NN$ interaction. We have demonstrated that the simple two-channel model including a single dibaryon pole in the internal channel is able to describe reasonably well the $NN$ scattering phase shifts and inelasticity parameters in all basic ($S$, $P$, $D$, $F$) partial waves in a broad energy range from zero up to 0.7--1.2 GeV using a few adjustable parameters in each partial wave. One should note in this respect that the high-precision $NN$ potentials based on the traditional meson-exchange picture as well as the state-of-the-art $NN$ potential derived within the chiral perturbation theory describe very well the elastic phase shifts up to 350 MeV and are not intended for reproducing the inelasticities. Though we employed the phenomenological determination of the model parameters in the present work, we have shown that these parameters can in principle be obtained from the six-quark microscopic treatment for the $NN$ system. An important result of the present study is that the dibaryon resonance parameters found from our model fit to the phase shifts turned out to be in good agreement with the empirical values.

Thus, we have found the direct connection between the
observable properties (masses, widths, decay modes) of dibaryons detected to date and
observables of $NN$ elastic and inelastic scattering. The next steps along
this way would be the calculations of the properties of few-body systems, finite nuclei and nuclear matter starting from the
$NN$ and $3N$ interactions derived within the dibaryon model.

We have also predicted a few dibaryon resonances in the partial waves where they have not been found yet.
This result stimulates further experimental research of $NN$ observables at intermediate energies. Thus, the new data on the sensitive spin-dependent observables could help refine the current $NN$ PWA solutions and reveal new dibaryon states. Still the main information on dibaryon resonances has been obtained from the inelastic processes involving the two-nucleon system. The experimental studies of photoproduction of mesons on the deuteron and deuteron photodisintegration recently undertaken and planned at, e.g., MAMI~\cite{Bash-em} and ELPH~\cite{Ishikawa19,Ishikawa21} seem to be very promising in exploring the properties of the known dibaryons as well as in finding the new ones. In view of the recent experimental progress in the field, one can say that we stand at the beginning of dibaryon spectroscopy.

The results of the present work support our claim that the six-quark states are not just exotic hadrons like tetra- or pentaquarks, but a regular mode in the fundamental $NN$ interaction carrying the basic intermediate-range attraction and also responsible for the short-range repulsion between nucleons. 
Thus, the dibaryon model can be considered as a new QCD-motivated tool appropriate for an effective treatment of the short-range nuclear force since it is free of some limitations of both quark and meson-exchange models and therefore has a wider range of applicability. These findings make further research of dibaryon resonances very perspective. It would be also interesting to study the connection between the $s$-channel mechanism suggested in this work and the contact terms used in the chiral perturbation theory.


\appendix

\section{Parametrization of the transition amplitude on the basis of the dressed bag model}
\label{formfactor}
In the dressed bag model~\cite{JPhys2001,IntJModPhys2002}, the transition from the internal ($6q+\sigma$)
channel to the external ($NN$) channel is described in terms of the RGM ansatz (\ref{RGM}).
In line with Eq.~(\ref{hrgm}), the transition matrix element can be written as
\begin{equation}
\begin{split}
h^{ex,in}|\alpha,{\bf k}\rangle\to \,
&\langle \{NN\}^Q|{\cal A}h_{6q}^{ex,in}|\alpha,{\bf k}\rangle\\
&=\langle \{NN\}^Q|\hat O_\sigma |\alpha,{\bf k}\rangle
\end{split}
\label{a1}
\end{equation}
where $[NN]^Q$ is an external $NN$ state orthogonalized to the $6q$ configuration
$s^6[6]_X$ (for a positive parity) or $s^5p[51]_X$ (for a negative parity):
\begin{equation}
|\{\!NN\!\}^{\!Q}\rangle\!=\!\Gamma^Q|\{\!NN\!\}\!\rangle,
\label{a2}
\end{equation}
where
\begin{equation}
\Gamma^Q\!=\!
\left \{\!\begin{array}{l}\!\! I\!-\!|\Psi_0\rangle\langle\Psi_0|,\,\, \mbox{for}\,S\,\,\mbox{wave},\\
\!\!I\!-\!|\Psi_1\rangle\langle\Psi_1|,\,\, \mbox{for}\, P\,\,\mbox{wave}, \end{array}\right\}
\label{a2a}
\end{equation}
 and $\hat O_\sigma$ is some operator of the $\sigma$-meson annihilation
(vertex $B$ in Fig.~\ref{f1}). Note that the operator of antisymmetrization
${\cal A}\!=\frac{1}{10}(I\!-\!\sum_{i\!=\!1}^3\!\sum_{j\!=\!4}^6P_{ij})$ is omitted in
the r.h.s. of Eq.~(\ref{a1}), since, by definition, both the $s^6$ and $s^5p$ components of the
dressed bag (i.e., the configurations $\Psi_0$ and $\Psi_1$) are antisymmetric under quark
permutations [see, e.g., Eq.~(\ref{apsi})].

It is important that the projection operator $\Gamma^Q$ defined in Eq.~(\ref{a2a}) may be
represented in terms of the excited $6q$ configurations $s^4p^2$ or $s^3p^3$, i.e., in terms of
the wavefunctions $\Psi^{(i)}_n$ defined by Eqs.~(\ref{q3}), (\ref{q3prime}) or
(\ref{q5}), (\ref{q5prime}), respectively. All one has to do is to construct the cluster
$3q+3q$ state with the quantum numbers of the $NN$ channel on the basis of the
shell-model wavefunctions $\Psi^{(i)}_n$:
\begin{equation}
 |Q_n^{NN}\rangle=
\sum_{i}U_{ni}^{NN}|\Psi_n^{(i)}\rangle.
\label{a3}
\end{equation}
Here $Q_n^{NN}({\bf r},\{{\bm\rho\xi}\})$ denotes the wavefunction of the cluster ($3q\!+\!3q$)
state at short distances and $U_{ni}^{NN}$ are the fractional parentage coefficients
(f.p.c.)~\cite{Harvey1979,Obukhovsky1996} for separation of two nucleonic $3q$ clusters
in the $6q$ configurations $\Psi_n^{(i)}$. For example, the $NN$ cluster state in the configuration
$s^4p^2$ with $ST\!=\!10$ (the $^3S_1$ partial wave) looks like:
\begin{equation}
\begin{split}
&|Q_2^{NN},ST\!=\!10\rangle\!=\!\\
&-\sqrt{\frac{9}{20}}\Psi_2^{(1)}\!+\!\sqrt{\frac{16}{45}}\Psi_2^{(2)}\!+\!
\sqrt{\frac{1}{36}}\Psi_2^{(3)}\!-\!\sqrt{\frac{1}{18}}\Psi_2^{(4)},
\end{split}
\label{a4}
\end{equation}
where the coefficients are from the first row of Tab.~\ref{TabA}, which lists the $6q\to3q+3q$ f.p.c.'s
for the configuration $s^4p^2[42]_X(ST\!=\!10)$.
The values of $U_{ni}^{NN}$ are defined as the amplitudes for the overlapping of the $CST$ parts of the $6q$ and
$3q\!+\!3q$ wavefunctions:
$$
U_{ni}^{NN}(CST)\!=\!\langle N(C_1S_1T_1\!)N(C_2S_2T_2)|\Psi_n^{(i)}\!(CST\!)\rangle,
$$
where $|N(CST)\rangle$ is given in Eq.~(\ref{q1}) and $|\Psi_n^{(i)}(CST)\rangle$ can be
taken from Eq.~(\ref{q3}). A special advantage of using the coefficients $U_{ni}^{NN}(CST)$ is that
they are independent of dynamics and are calculated by the group theoretical methods~\cite{Obukhovsky1979,Obukhovsky1996}.

With the cluster $NN$ state defined by Eq.~(\ref{a3}), one may represent the projection operator (\ref{a2}) in the form
\begin{equation}
\Gamma^Q= |Q_n^{NN}\rangle\langle Q_n^{NN}|+\dots,
\label{a5}
\end{equation}
where dots symbolize terms of little importance at short range. Then the transition matrix element (\ref{a1}) takes the form:
\begin{equation}
\begin{split}
&h^{ex,in}|\alpha,{\bf k}\rangle=\langle \{NN\}^Q|\hat O_\sigma |\alpha,{\bf k}\rangle=\\
&\langle \{NN\}|\sum_iU_{ni}^{NN}|\Psi_n^{(i)}\rangle\sum_j U_{nj}^{NN}
\langle\Psi_n^{(j)}|\hat O_\sigma|\alpha,{\bf k}\rangle,
\end{split}
\label{a6}
\end{equation}
and thus, both the form factor $Z^{JM}_{LS}(r)$ and vertex function
$B_{L_{\sigma}LS}^J(k,E)$ introduced in Eq.~(\ref{tr2}) (see Sec.~\ref{pole})
obtain a microscopic interpretation in terms of Eq.~(\ref{a6}):
\begin{equation}
Z_n({\bf r})=\sum_iU_{ni}^{NN}\langle \{NN\}_{ST}|\Psi_n^{(i)}\rangle,
\label{a7}
\end{equation}
\begin{equation}
B_n({\bf k},E)=\sum_j U_{nj}^{NN}\langle\Psi_n^{(j)}|
\hat O_\sigma({\bf k},E)|\alpha,{\bf k}\rangle.
\label{a8}
\end{equation}
After a partial-wave decomposition of the r.h.s. of Eqs.~(\ref{a6})--(\ref{a8}), the functions $Z_n$
and $B_n$ transform into $Z^{JM}_{LS}(r)$ and $B_{L_{\sigma}LS}^J(k,E)$, respectively, and
Eq.~(\ref{a6}) becomes equivalent to Eq.~(\ref{tr2}).

It is important that the quark model restricts the form of the overlap functions
$\langle \{NN\}|\Psi_n^{(i)}\rangle$ in the r.h.s. of Eq.~(\ref{a7}), as is seen
from the comparison of these functions with those defined in Eqs.~(\ref{q9}) and (\ref{q15}).
For example, keeping in mind that at $n=$2(3) the values of $LST$ should be fixed at
$L\!=\!$0(1), $ST$=10,01(11,00), one obtains the following expression for the partial
wave decomposition of the form factor~(\ref{a7}):
\begin{equation}
\begin{split}
Z^{JM}_{LS}(r)\!&=\!\sum_i\!U_{ni}^{N\!N}\!\left\{\!\!\int \!\!Y_L^*(\hat r)
d\Omega_{r}\langle \{\!NN\!\}_{ST}|\Psi_n^{(i)}\rangle\!\right\}_{\!\!\!JM}\\
&=\left \{\!\begin{array}{l}{\cal N}_n(1\!-\!\frac{r^2}{b^2})\exp(-\frac{3r^2}{4b^2}),\,n\!=\!2\,,\\
{\cal N}_nr(1\!-\!\frac{r^2}{b^2})\exp(-\frac{3r^2}{4b^2}),\,n\!=\!3\,, \end{array}\right.
\end{split}
\label{a9}
\end{equation}
where the normalization factors ${\cal N}_n,\,n\!=\!2,3$ should take into account the contributions
of both the coefficients $U_{ni}^{N\!N}$ and the overlap integrals over the inner nucleonic
coordinates $\{{\bm \rho\xi}\}$. Note that the integration over $\Omega_r$ in Eq.~(\ref{a9})
is formal here, since the value of $L$ is already fixed by the given value of $n$.

The quark model also restricts the form of the vertex functions
$B_{L_{\sigma}LS}^J(k)$. As is evident from Eq.~(\ref{a8}), the functions $B_n({\bf k},E)$
should be proportional to the shell-model matrix elements
$\langle\Psi_n^{(j)}|\hat O_\sigma({\bf k},E)|\Psi_m\rangle,\,m=0,1$, where $\Psi_m$ means
the $6q$ component of the dressed bag state
$|\alpha,{\bf k}\rangle$~\cite{JPhys2001,IntJModPhys2002}.
The corresponding wavefunctions are given in Eqs.~(\ref{q2})--(\ref{q2prime}) (for a positive parity)
and (\ref{q4})--(\ref{q4prime}) (for a negative parity). The partial wave decomposition of $B_n({\bf k},E)$
is analogues to Eq.~(\ref{a9}), corrected for the substitution of
$\langle\Psi_n^{(j)}|\hat O_\sigma({\bf k},E)|\Psi_m\rangle$ for
$\langle \{\!NN\!\}_{ST}|\Psi_n^{(i)}\rangle$ (see Appendix~\ref{dressing} for the calculation of the
shell-model matrix element $\langle\Psi_n^{(j)}|\hat O_\sigma({\bf k},E)|\Psi_m\rangle$).

\begin{table}[h!]
\caption{\label{TabA} Baryon-baryon ($B_1B_2$) content of the $6q$ configuration
$s^4p^2[42]_X[f_i]_{CS}ST\!=\!10$ ($i=\,$0,1,\dots5) represented by the f.p.c.'s $U^{B_1B_2}_{2i}$. The notation
$U_{\!2i}([42]_{\scriptscriptstyle X}\!)$ is used for the coefficients $U^{B_1B_2}_{2i}$
which appear in Eqs.~(\ref{a3})--(\ref{a6}) for $B_1B_2\!=\!NN$.}
\begin{center}
\begin{tabular}{|c|c|c|c|c|c|c|}
\hline
&\multicolumn{6}{l|}
{$U_{\!20}([6]_{\!\scriptscriptstyle X}\!\!)|$\qquad\qquad\qquad
$U_{\!2i}([42]_{\scriptscriptstyle X}\!)$}\\
\cline{2-7}
\raisebox{1.5ex}[0cm][0cm]{}
&$\,\,[2^3]_{CS}\,\,$&$[42]_{CS}$&$[321]_{CS}$&$[2^3]_{CS}$&$[31^3]_{CS}$&$[21^4]_{CS}$ \\
\raisebox{0.6ex}[0cm][0cm]{$B_1B_2\backslash i$} & 0 & 1 & 2 & 3 & 4 & 5 \\
\hline
$NN$ &$\sqrt{\frac{1}{9}}$ &-$\sqrt{\frac{9}{20}}$ &$\sqrt{\frac{16}{45}}$
& $\sqrt{\frac{1}{36}}$&-$\sqrt{\frac{1}{18}}$& 0 \\
\hline
$\Delta\Delta$ &-$\sqrt{\frac{4}{45}}$& 0 & 0 & $\sqrt{\frac{16}{45}}$& 0 & $\sqrt{\frac{5}{9}}$\\
\hline
$C_1C_1$ &$\sqrt{\frac{2}{9}}$&-$\sqrt{\frac{1}{10}}$&-$\sqrt{\frac{8}{45}}$&
 $\sqrt{\frac{1}{18}}$&$\sqrt{\frac{4}{9}}$& 0\\
\hline
$C_1C_2$ &$\sqrt{\frac{4}{9}}$&$\sqrt{\frac{1}{5}}$&-$\sqrt{\frac{1}{45}}$&
$\sqrt{\frac{1}{9}}$&-$\sqrt{\frac{2}{9}}$& 0\\
\hline
$C_2C_2$ &$\sqrt{\frac{1}{45}}$&$\sqrt{\frac{1}{4}}$&$\sqrt{\frac{4}{9}}$&
$\sqrt{\frac{1}{180}}$&$\sqrt{\frac{5}{18}}$& 0\\
\hline
$C_3C_3$ &-$\sqrt{\frac{1}{9}}$&0&0&$\sqrt{\frac{4}{9}}$&0&-$\sqrt{\frac{4}{9}}$\\
\hline
\end{tabular}
\end{center}
\end{table}

\section{Quark-model calculation of the transition amplitude}
\label{dressing}

\begin{figure}[h]
\begin{center}
 \epsfig{file=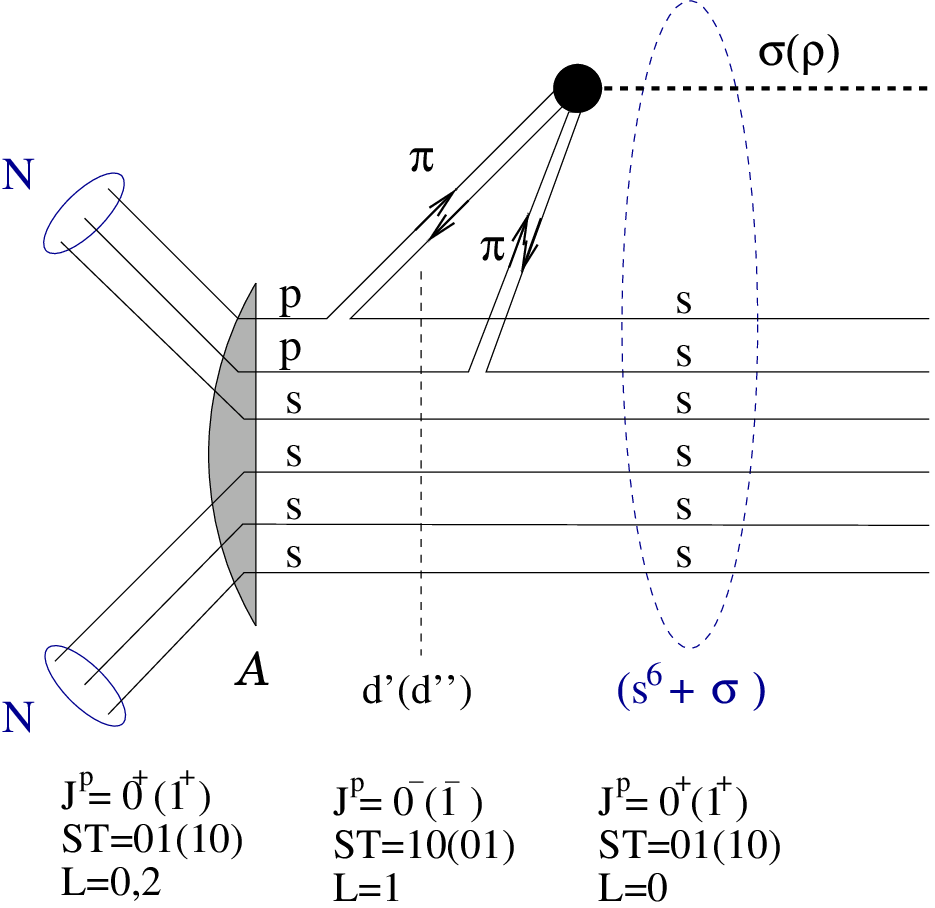,width=0.75\columnwidth}
\end{center}
\caption{The transition from the $NN$ to the dressed bag channel via
two sequential pion emissions from two $p$-shell quarks.\label{figB1}}
\end{figure}

The vertex functions $B_{L_{\sigma}LS}^J(k,E)$ determining the transition
from the external $NN$ to the internal $(6q+\sigma)$ channel were calculated in
Refs.~\cite{JPhys2001,IntJModPhys2002} for the lowest even partial waves (L=0,2), assuming the
emission of the $s$-wave $\sigma$ meson ($L_{\sigma}\!\!=0$) from the initial $|s^4p^2[42]_X\rangle$
six-quark state. The calculations were performed
within the framework of TISM including all
$6q$ configurations with 0$\hbar\omega$, 1$\hbar\omega$ and 2$\hbar\omega$ excitations,
subsequently denoted as $d_0$, $d_1^\prime(d_1^{\prime\prime})$ and $d_2$.
For this, the two-pion emission amplitude in the transition from the orthogonalized [in a sense of
Eq.~(\ref{orth})] $NN$ state to the dressed $6q$ bag $d_0+\sigma(2\pi)$ was calculated.
The graph illustrating this process is shown in Fig.~\ref{figB1}.

The transition goes in two steps via intermediate $6q$ states $d_1'(d_1'')$:
\begin{equation}
[NN]^Q\to d_1'(d_1'')+\pi\to d_0+\sigma(2\pi)
\label{nnq}
\end{equation}
(for the even
partial waves $L$=0, 2 in the initial $NN$ state).
The final $6q$ state in Fig.~\ref{figB1} is the most symmetric (and compact)
shell-model quark configuration $d_0=s^6[6]_X$. The intermediate
six-quark states $d_1^{\prime}$($d_1^{\prime\prime}$) denoted by the vertical
dashed line in the graph belong to the configurations $s^5p[51]_X[f]_{CS}(ST)$
and have the following quantum
numbers: $d_1'(ST\!=\!10,J^P\!=\!0^-)$
for the singlet $NN$ channel\footnote{These are the quantum numbers of the
so-called $d^{\prime}$-dibaryon~\cite{dprime}}
and $d_1''(ST\!=\!01,J^P\!=\!1^-)$ for the triplet channel.

\begin{figure}[h]
\begin{center}
 \epsfig{file=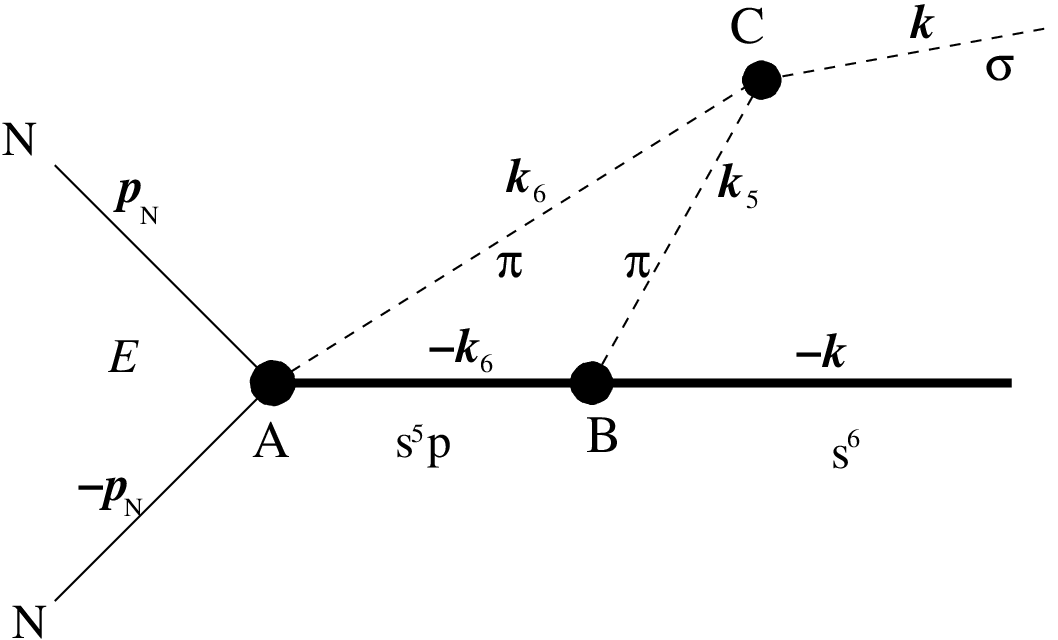,width=0.85\columnwidth}
\end{center}
\caption{\label{figB2} Triangle diagram corresponding to
the $\sigma$-meson formation from two pions in the transition
of two $p$-shell quarks to the $s$ orbit ($k_5$ and $k_6$ are the momenta of the 5-th and 6-th
quarks in the diagram of Fig.~\ref{figB1}).}
\end{figure}

Here we briefly outline the scheme for calculating the
amplitude~(\ref{nnq}) with the successive emission of two pions shown in Fig.~\ref{figB1}. The amplitude
can be represented in a simplified form as a triangular diagram that does not explicitly
contain the quark lines (see Fig.~\ref{figB2}).

As can be seen from these diagrams, to calculate the respective amplitude, it is necessary to determine
three vertices ($A$, $B$ and $C$) and carry out the integration over the momenta ${\bf k}_5$ and ${\bf k}_6$
of the intermediate pions emitted by the 5-th and 6-th quarks.
The initial six-quark $NN$ state is written in a form of the RGM ansatz (\ref{RGM}) and decomposed into
the basis of the TISM $6q$ configurations with two quanta of excitation (see Appendix~\ref{formfactor}
for detail). This decomposition includes configurations $s^4p^2$, $s^52s$, and $s^5d$.

At the vertex $A$, a transition occurs from these initial configurations to an intermediate state $s^5p$ with the
emission of a pion by one of the $p$-shell quarks: $s^4p^2 (s^52s\,\, \mbox{or}\,\,s^5d)\to s^5p+\pi$.
At the vertex $B$, the $p$-shell quark of the $s^5p$ configuration passes into the $s$-state with the emission
of a second pion: $s^5p\to s^6+\pi$. The pion creation amplitudes in these vertices were calculated within
the framework of the well-known quark-pair-creation model~\cite{book,Desp,Karmanov} with using the
TISM matrix elements for the $6q$ transitions.

The $\pi +\pi \to \sigma$ transition amplitude (the vertex $C$ in Fig.~\ref{figB2}) is
 to be proportional to the overlap of the two-pion
and the $\sigma$-meson wavefunctions~\cite{Guts94}:
\begin{align}
\langle \pi({\bf k})\pi({\bf k}^\prime)|H_{\pi\pi\sigma}|\,\sigma\rangle
=f_{\pi\pi\sigma}F_{\pi\pi\sigma}({\bm\varkappa}^2),\nonumber \\
F({\bm \varkappa}^2)=\exp(-\frac{1}{2}\varkappa^2b_{\sigma}^2),
\label{sigm}
\end{align}
where ${\bm \varkappa}=\frac{1}{2}({\bf k}-{\bf k}^\prime)$
and $b_{\sigma}$ is a characteristic scale of the $\sigma$ meson in the
$\pi\pi$ channel.

In the calculations, the shell-model quark representations for the
pion and the $\sigma$ meson were used~\cite{Guts94}:
\begin{eqnarray}
|\pi^{\lambda}\rangle =\vert s\bar s[2]_XLST=001\,T_z\mbox{=}\lambda\,\,
J^P\!\!=0^-\rangle, \nonumber\\
 |\sigma\rangle =|s^2\bar{s}^2[4]_X,LST\mbox{=}000,J^P=0^+\rangle ,
 \label{qconf}
\end{eqnarray}
with the Gaussian wavefunctions, e.g., $\Psi_{\pi}({\bf \rho_{\pi}})\sim
\exp(-\rho_{\pi}^{2}/4b_{\pi}^2),\quad {\bf \rho}_{\pi}(ij)={\bf r}_i-{\bf
r}_j$, where $b_{\pi}$ is the ``quark radius'' of the pion ($\approx$ 0.5$b
\approx$0.3~fm).

When calculating the triangle diagram, three possible temporal orderings of the vertices were taken
into account: $ABC$, $ACB$, $CBA$. As a result of integration over the inner pion momentum, the explicit
expressions for the vertex functions $B_{L_\sigma LS}^J(k,E)$ were obtained~\cite{JPhys2001,IntJModPhys2002}:
\begin{equation}
B^J_{L_\sigma LS}(k,E)=g^J_{L_\sigma LS}D^J_{L_\sigma LS}(k,E).
\label{BgD}
\end{equation}
Here the factor $g^J_{L_\sigma LS}$ can be considered as an effective coupling constant with the
value $g^J_{0 LS}=\frac{f^2_{\pi qq}}{m^2_\pi}\frac{g_{\sigma \pi\pi}}{m^2_qb^3}C^J_{LS}$ defined
in~\cite{IntJModPhys2002} at $L_\sigma=0$. The standard coupling constants were used for
the $\pi qq$ and $\sigma\pi\pi$ vertices: $f_{\pi qq}=\frac{3}{5}f_{\pi NN}$ and
$g_{\sigma\pi\pi}\!\approx$2-4 $GeV/c$, while the parameters $C^J_{LS}$ were calculated by the
f.p.c. technique. The vertex functions $D^J_{0 LS}(k,E)$ are rather cumbersome
expressions that include the integration over the inner momenta of the diagram in Fig.~\ref{figB2}.
We do not present these vertex functions here, since the definition of the effective potential $w(E)$
for the external channel includes only their convolutions with the meson propagator, which determine
the energy-dependent coupling constants $\lambda_{LL'}^J(E)$ (\ref{lamb}). To approximately
reproduce the energy dependence of the constants $\lambda_{LL'}^J(E)$ derived from the
microscopic calculation, the Pade approximant [1,1] was used in Refs.~\cite{JPhys2001,IntJModPhys2002}
and in subsequent applications of the dibaryon model.



\begin{thebibliography}{abc}
\bibitem{Q1} S. Aoki and T. Doi, Front. Phys. {\bf 8}, 307 (2020).
\bibitem{Q2} F. Fernandez, P.G. Ortega, and D.R. Entem, Front. Phys. {\bf 7}, 233 (2020).
\bibitem{Kuk-YAF11} V.I. Kukulin, Phys. Atom. Nucl. {\bf 74}, 1567 (2011).
\bibitem{Clem17} H. Clement, Prog. Part. Nucl. Phys. {\bf 93}, 195 (2017).
\bibitem{Clem21} H. Clement and T. Skorodko, Chin. Phys. C {\bf 45} 2, 022001 (2021).
\bibitem{Gal16} A. Gal, Acta Phys. Polon. B {\bf 47}, 471 (2016).
\bibitem{PIYAF} V.I. Kukulin, in: Proceedings of XXXIII Winter
School PIYaF, Gatchina, 1999, p. 207.
\bibitem{JPhys2001}  V.I. Kukulin, I.T. Obukhovsky, V.N. Pomerantsev, and
A. Faessler, J. Phys. G {\bf 27}, 1851 (2001).
\bibitem{IntJModPhys2002} V.I. Kukulin, I.T. Obukhovsky, V.N. Pomerantsev, and
A. Faessler, Int. J. Mod. Phys. E {\bf 11}, 1 (2002).
\bibitem{dual} M.I. Krivoruchenko and A. Faessler, Rom. J. Phys. {\bf 57}, 296 (2012).
\bibitem{PAN13} V.I. Kukulin and M.N. Platonova, Phys. At. Nucl. {\bf 76}, 1465
(2013).
\bibitem{AnnPhys2010} V.I. Kukulin {\it et al.}, Ann. Phys. (NY) {\bf 325}, 173 (2010).
\bibitem{YAF19}
V.I. Kukulin, V.N. Pomerantsev, O.A. Rubtsova, and M.N. Platonova, Phys. At. Nucl. {\bf 82}, 934
(2019).
\bibitem{PLB20} V.I. Kukulin {\it et al.}, Phys. Lett. B {\bf 801}, 135146 (2020).
\bibitem{EPJA20} V.I. Kukulin {\it et al.}, Eur. Phys. J. A {\bf 56}, 229 (2020).
\bibitem{PRD20} O.A. Rubtsova, V.I. Kukulin, and M.N. Platonova,
Phys. Rev. D {\bf 102}, 114040 (2020).
\bibitem{Dyson64} F.J. Dyson and N.-H. Xuong, Phys. Rev. Lett. {\bf 13}, 815 (1964); Erratum {\em ibid.} {\bf 14}, 339 (1965).
\bibitem{Meshcher} M.G. Meshcheriakov, B.S. Neganov, Dokl. Akad. Nauk SSSR {\bf 100}, 677 (1955);
B.S. Neganov, L.B. Parfenov, Soviet Phys. JETP {\bf 7}, 528
(1958).
\bibitem{Hoshiz} N. Hoshizaki, Prog. Theor. Phys. {\bf 60}, 1796 (1978);
\emph{ibid.} {\bf 61}, 129 (1979); \emph{ibid.} {\bf 89}, 251 (1993); \emph{ibid.} {\bf 89}, 569 (1993).
\bibitem{Arndt} R. Bhandari, R.A. Arndt, L.D. Roper, B.J. VerWest, Phys. Rev. Lett. {\bf 46}, 1111
(1981); C.-H. Oh, R.A. Arndt, I.I. Strakovsky, R.L. Workman,
Phys. Rev. C {\bf 56}, 635 (1997).
\bibitem{Strak} A.V. Kravtsov, M.G. Ryskin, I.I. Strakovsky, J. Phys. G {\bf 9},
L187 (1983); I.I. Strakovsky, A.V. Kravtsov, M.G. Ryskin, Sov. J.
Nucl. Phys. {\bf 40}, 273 (1984).
\bibitem{Gal14} A. Gal and H. Garcilazo, Nucl. Phys. {\bf A928}, 73 (2014); Phys. Rev. Lett. {\bf 111}, 172301 (2013).
\bibitem{Kamae77} T. Kamae \emph{et al.}, Phys. Rev. Lett. {\bf 38}, 468 (1977); T. Kamae and T. Fujita, \emph{ibid.} {\bf 38}, 471 (1977).
\bibitem{Bash09} M. Bashkanov {\em et al.} (CELSIUS/WASA Collaboration), Phys. Rev. Lett. {\bf 102}, 052301 (2009).
\bibitem{Adl1113} P. Adlarson {\em et al.} (WASA-at-COSY Collaboration), Phys. Rev. Lett. {\bf 106}, 242302 (2011); Phys. Lett. B {\bf 721}, 229 (2013).
\bibitem{AD14} P. Adlarson {\em et al.} (WASA-at-COSY Collaboration and SAID Data Analysis Center), Phys. Rev. Lett. {\bf 112},
202301 (2014); Phys. Rev. C {\bf 90}, 035204 (2014); {\em ibid.} {\bf 102}, 015204 (2020).
\bibitem{Bash-em} M. Bashkanov {\em et al.} (A2 Collaboration), Phys. Rev. Lett. {\bf 124}, 132001 (2020); M. Bashkanov {\em et al.}, Phys. Lett. B {\bf 789}, 7 (2019) and references therein.
\bibitem{BBC13} M. Bashkanov, S.J. Brodsky, and H. Clement, Phys. Lett. B {\bf 727}, 438 (2013).
\bibitem{Huang} F. Huang, Z.Y. Zhang, P.N. Shen, and W.L. Wang, Chin. Phys. C {\bf 39}, 071001 (2015); F. Huang, P.N. Shen, Y.B. Dong, and Z.Y. Zhang, Sci. China Phys. Mech. Astron. {\bf 59}, 622002 (2016).
\bibitem{Nisk17} J.A. Niskanen, Phys. Rev. C {\bf 95}, 054002 (2017).
\bibitem{Adl18d21} P. Adlarson {\em et al.} (WASA-at-COSY Collaboration), Phys. Rev. Lett. {\bf 121}, 052001 (2018); 
Phys. Rev. C {\bf 99}, 025201 (2019).
\bibitem{Adl16d30} P. Adlarson {\em et al.} (WASA-at-COSY Collaboration), Phys. Lett. B {\bf 762}, 455 (2016).
\bibitem{Auer1} I.P. Auer et al., Phys. Lett. {\bf 67B}, 113 (1977); I.P. Auer et al., Phys. Lett.
{\bf 70B}, 475 (1977); K. Hidaka et al., Phys. Lett. {\bf 70B},
479 (1977).
\bibitem{Auer2} I.P. Auer et al., Phys. Rev. Lett. {\bf
41}, 354 (1978); I.P. Auer et al., Phys. Rev. Lett. {\bf 41}, 1436
(1978); I.P. Auer et al., Phys. Rev. Lett. {\bf 48}, 1150 (1982).
\bibitem{MacGregor} M.H. MacGregor, Phys. Rev. D {\bf 20}, 1616 (1979).
\bibitem{Nijm} P.J. Mulders, A.T.M. Aerts, and J.J. De Swart, Phys. Rev.
D {\bf 21}, 2653 (1980).
\bibitem{ITEP} L.A. Kondratyuk, B.V. Martemyanov, M.G. Shchepkin,
Sov. J. Nucl. Phys. {\bf 45}, 776 (1987).
\bibitem{dprime} W. Brodowski \emph{et al.}, Z. Phys. A {\bf 355}, 5 (1996).
\bibitem{dprime1}
W. Brodowski \emph{et al.}, Phys. Lett. B {\bf 550}, 147 (2002).
\bibitem{SAID} R.L. Workman, W.J. Briscoe, and I.I. Strakovsky, Phys. Rev. C {\bf 94}, 065203 (2016);
all SAID PWA solutions can be accessed via the website: {http://gwdac.phys.gwu.edu}
\bibitem{Komarov16} V. Komarov \emph{et al.}, Phys. Rev. C {\bf 93}, 065206 (2016).
\bibitem{NPA2016} M.N. Platonova and V.I. Kukulin, Nucl. Phys. {\bf A946}, 117 (2016).
\bibitem{PRD2016} M.N. Platonova and V.I. Kukulin, Phys. Rev. D {\bf 94}, 054039 (2016).
\bibitem{Strak91}
I.I. Strakovsky, Fiz. Elem. Chast. Atom. Yadra {\bf 22}, 615
(1991); AIP Conf. Proc. {\bf 221}, 218 (1991).
\bibitem{Nisk20} J.A. Niskanen, Phys. Rev. C {\bf 102}, 024002 (2020).
\bibitem{Clem20} H. Clement, T. Skorodko, and E. Doroshkevich, Phys. Rev. C (in press); arXiv:2010.09217 [nucl-ex].
\bibitem{Ishikawa19} T. Ishikawa \emph{et al.}, Phys. Lett. B {\bf 789}, 413 (2019); Y. Toyama \emph{et al.} (NKS2
Collaboration), Few-Body Syst. {\bf 63}, 15 (2022).
\bibitem{Jude22} T.C. Jude \emph{et al.}, Phys. Lett. B {\bf 832}, 137277 (2022).
\bibitem{Ishikawa21} T. Ishikawa \emph{et al.}, Phys. Rev. C {\bf 104}, L052201 (2021).
\bibitem{Tsirkov19} D. Tsirkov \emph{et al.}, EPJ Web Conf. {\bf 199}, 02016 (2019).
\bibitem{Tsirkov22} D. Tsirkov \emph{et al.}, arXiv:2207.13575 [nucl-ex].
\bibitem{Alde97}
D. Alde {\em et al.} (GAMS Collaboration), Phys. Lett. B {\bf 397}, 350 (1997).
\bibitem{PRC2013} M.N. Platonova and V.I. Kukuin, Phys. Rev. C {\bf 87}, 025202 (2013).
\bibitem{PRD2021} M.N. Platonova and V.I. Kukulin, Phys. Rev. D {\bf 103}, 114025 (2021).
\bibitem{Eichten} E. Eichten, S. Godfrey, H. Mahlke, and J.L. Rosner,
Rev. Mod. Phys. {\bf 80}, 1161 (2008).
\bibitem{Kusainov1991}
A.M. Kusainov, V.G. Neudatchin, and I.T. Obukhovsky, Phys. Rev. C {\bf 44}, 2343 (1991);
I.T. Obukhovsky and A.M. Kusainov, Phys. Lett. B {\bf 238}, 142 (1990).
\bibitem{PDG}
P.A. Zyla {\em et al.} (Particle Data Group), Prog. Theor. Exp.
Phys. {\bf 2020}, 083C01 (2020).
\bibitem{Glozman}
L.Ya. Glozman, Phys. Lett. {\bf B475}, 329 (2000); L.Ya. Glozman and A.V. Nefediev, Phys. Rev. D {\bf 76},
096004 (2007).
\bibitem{Volkov}
M.K. Volkov, A.E. Radzhabov, and N.L. Russakovich, Phys.
At. Nucl. {\bf 66}, 997 (2003); M.K. Volkov \emph{et al.}, Phys. Lett. {\bf B424},
235 (1998).
\bibitem{Saito1969} S. Saito, Prog. Theor. Phys. {\bf 41}, 705 (1969).
\bibitem{Obukhovsky1996}
I.T. Obukhovsky, Prog. Part. Nucl. Phys. {\bf 36}, 359 (1996).
\bibitem{Harvey1979}
M. Harvey, Nucl. Phys. A {\bf 352}, 301 (1981); A {\bf 352}, 326 (1981).
\bibitem{Obukhovsky1979}
I.T. Obukhovsky, V.G. Neudatchin, Yu.F. Smirnov, and Yu.M. Tchuvil'sky, Phys. Lett. {\bf 88}B,
231 (1979).
\bibitem{DeRujula1975}
A. De Rujula, H. Georgi, and S.L. Glashow, Phys. Rev. D {\bf 12}, 147 (1975).
\bibitem{Neudatchin1975} V.G. Neudatchin, I.T. Obukhovsky, V.I. Kukulin, and N.F. Golovanova,
Phys. Rev. C {\bf 11}, 128 (1975).
\bibitem{Neudatchin1977} V.G. Neudatchin, Yu.F. Smirnov, and R. Tamagaki,
Prog. Theor. Phys. {\bf 58}, 1072 (1977).
\bibitem{Sazonov} V.M. Krasnopolsky, V.I. Kukulin, V.N. Pomerantsev, P.B. Sazonov,
Phys. Lett. {\bf B165}, 7 (1985).
\bibitem{sys3n} V.I. Kukulin, V.N. Pomerantsev, M. Kaskulov, and
A. Faessler, J.~Phys.~G {\bf 30}, 287 (2004);
V.I.~Kukulin, V.N.~Pomerantsev, and A.~Faessler, J.~Phys.~G {\bf 30}, 309 (2004).
\bibitem{YAF3N} V.N. Pomerantsev, V.I. Kukulin, V.T. Voronchev, and A. Faessler, Phys. Atom. Nucl. {\bf 68}, 1453 (2005).
\bibitem{6Li} M. Kakenov, V.I. Kukulin, V.N. Pomerantsev, O. Bayakhmetov,
 Eur. Phys. J. A {\bf 56}, 266 (2020).
\bibitem{Obukh19}
I.T. Obukhovsky, A. Faessler, D.K. Fedorov, T. Gutsche, and V.E. Lyubovitskij,
Phys. Rev. D {\bf 84}, 014004 (2011); Phys. Rev. D {\bf 100}, 014032 (2019).
\bibitem{book} A. Le Yaouanc, L. Oliver, O. P{\' e}ne, and J.-C. Raynal, {\it Hadron Transitions
in the Quark Model} (Gordon and Beach Science Publishers, NY), 1988.
\bibitem{Desp} F. Cano, F. Gonz{\' a}lez, S. Noguera, and B. Deplanques, Nucl. Phys. {\bf A603},
257 (1996).
\bibitem{Karmanov} J. Carbonell, B. Deplanques, V.A. Karmanov, and J.F. Mathiot, Phys. Rep. {\bf 300},
215 (1998).
\bibitem{Guts94} T. Gutsche, R.D. Viollier, and A. Faessler, Phys. Lett. B {\bf 331}, 8 (1994).

\end{thebibliography}
\end{document}